\documentclass[review]{elsarticle}

\usepackage{lineno,hyperref}
\modulolinenumbers[5]

\journal{.....}

\usepackage{natbib}
\usepackage{amsmath}
\usepackage{amsfonts}
\usepackage{relsize}
\usepackage{appendix}
\usepackage{amsthm}
\usepackage{color}
\usepackage{afterpage}

\usepackage[normalem]{ulem}

\newtheorem{thm}{Theorem}
\newtheorem{lem}[thm]{Lemma}
\newtheorem{pf*}{Proof}

\biboptions{authoryear}

\begin{document}

\begin{frontmatter}

\title{Nonparametric incidence estimation and bootstrap bandwidth selection in mixture cure models}

\author[mymainaddress1]{Ana L\'{o}pez-Cheda\corref{mycorrespondingauthor}}
\ead{ana.lopez.cheda@udc.es}
\cortext[mycorrespondingauthor]{Correspondence to: Departamento de Matem\'{a}ticas, Facultade de Inform\'{a}tica, Universidade da Coru\~{n}a, A Coru\~{n}a 15071, Spain. Tel.: +34 981 16 70 00 Ext. 1301; fax: +34 981 16 70 00}

\author[mymainaddress1]{Ricardo Cao}

\author[mymainaddress1]{Mar\'{i}a Amalia J\'{a}come}

\author[mymainaddress2]{Ingrid Van Keilegom}


\address[mymainaddress1]{Departamento de Matem\'{a}ticas, Universidade da Coru\~{n}a, Spain}
\address[mymainaddress2]{Institut de Statistique, Biostatistique et Sciences Actuarielles, Universit\'{e} catholique de Louvain, Belgium}

\begin{abstract}
A completely nonparametric method for the estimation of mixture cure mo\-dels is proposed. A nonparametric estimator of the incidence is extensively studied and a nonparametric estimator of the latency is presented. These estimators, which are based on the Beran estimator of the conditional survival function, are proved to be the local maximum likelihood estimators. An iid representation is obtained for the nonparametric incidence estimator. As a consequence, an asymptotically optimal bandwidth is found. Moreover, a bootstrap bandwidth selection method for the nonparametric incidence estimator is proposed. The introduced nonparametric estimators are compared with existing semiparametric approaches in a simulation study, in which the performance of the bootstrap bandwidth selector is also assessed. Finally, the method is applied to a database of colorectal cancer from the University Hospital of A Coru\~{n}a (CHUAC).
\end{abstract}


\end{frontmatter}

\linenumbers

\section{Introduction}
\label{sec:1} 

\nolinenumbers

Thanks to the effectiveness of current cancer treatments, the proportion of patients who get cured (or who at least survive for a long time) is increasing over time. Therefore, data coming from cancer studies typically have heavy censoring at the end of the follow-up period, and a standard survival model is inappropriate. To accommodate for the cured or insusceptible proportion of subjects, a cure fraction can be explicitly incorporated into survival models and, as a consequence, cure models arise. These models allow to estimate the cured proportion (incidence) and also the probability of survival of the uncured patients up to a given time point (latency). Note that cure models should not be used indiscriminately \citep{Farewell2}, there must be good empirical and biological evidence of an insusceptible population.

There are two main classes of cure models: mixture and non-mixture mo\-dels.
The first papers in non-mixture models were due to \cite{Haybittle1}, \cite{Haybittle2}. One category, belonging to this group, is the proportional hazards (PH) cure
model, also known as the promotion time cure model, first proposed by \cite{Yakovlev2}. The unknown terms in this model can be estimated parametrically \citep{Yakovlev1, Chen2, Chen1} or semiparametrically \citep{Tsodikov1, Tsodikov3, Zeng}. Moreover, \cite{Tsodikov2} proposed a nonparametric estimator of the incidence, but it cannot handle continuous covariates.

In this paper we consider a model which belongs to the other category of cure \mbox{models}, called two-component mixture cure models. The mixture cure model was proposed by \cite{Boag} and it explicitly expresses the survival function in terms of a mixture of the survival of two types of patients: those who are cured and those who are not. An advantage of this model is that it allows the covariates to have different influence on cured and uncured patients. \cite{Maller3} provided a detailed review of this model. In mixture cure models, the incidence is usually assumed to have a logistic form and the latency is usually estimated parametrically \citep{Farewell1, Farewell2, Cantor, Ghitany, Denham} or semiparametrically \citep{Kuk, Yamaguchi, Peng2, Peng1, Sy, Li1, Zhang}. Some recent topics covered in the mixture cure models literature include multivariate survival data \citep{Yu}, clustered survival data \citep{Lai2} and accelerated models \citep{Zhang2}.

Due to the fact that the effects of the covariate on the cure rate cannot
always be well appro\-xi\-ma\-ted using parametric or semiparametric
methods, a nonparametric approach is needed. In the literature, some
nonparametric methods for the estimation of the cure rate have been studied:
\cite{Maller1} proposed a consistent nonparametric estimator of the
incidence, but it cannot handle covariates. In order to overcome this
drawback, \cite{Laska} proposed another nonparametric estimator
of the cure rate, but it only works for discrete covariates. Furthermore,
\cite{Wang} proposed a cure model with a nonparametric form in the
cure probability. To ensure model \mbox{identifiability}, they assumed a
nonparametric proportional ha\-zards model for the hazard function. The
estimation was carried out by an expectation-maximization algorithm for a
penalized likelihood. They defined the smoothing spline func\-tion estimates
as the minimizers of the penalized likelihood. More recently, \cite{Xu} extended the existing work by proposing a nonparametric
incidence estimator which allows for a continuous covariate. Although the
above papers have a nonparametric flavor, they fail to consider a completely
nonparametric mixture cure model which works for discrete and con\-ti\-nuous
covariates in both the incidence and the latency. 

In this paper, we fill this important gap by proposing a two-component mixture model with non\-para\-metric forms for both the cure probability and the survival function of the uncured individuals. Although we consider only one covariate, the method can be directly extended to a case with multiple covariates.

Very recently \cite{Lopezetal}  have carried out a detailed study of the nonparametric kernel latency estimator proposed in this paper. They have proven asymptotic properties for the latency estimator and proposed a bandwidth selector.

The rest of the article is organized as follows. In Section \ref{sec:2} we
give a detailed description of our nonparametric mixture cure model, we
study the estimator of the incidence proposed in \cite{Xu} and we introduce a nonparametric estimator of the latency. Moreover, we address the model identifiability. We also present a local maximum likelihood result as well as an iid representation and the asymptotic mean squared error for the nonparametric incidence estimator. A bootstrap bandwidth selection
method is introduced in Section \ref{sec:3}. Section \ref{sec:4} includes a
comparison between these nonparametric estimators and the semiparametric
one proposed in \cite{Peng1} in a simulation study and assesses the
practical performance of the bootstrap bandwidth selector. In Section \ref{sec:5} we apply the proposed nonparametric method to real data related to colorectal cancer patients in CHUAC. An appendix contains the proofs.

\section{Nonparametric mixture cure model}
\label{sec:2}
\subsection{Notation}

\nolinenumbers

Let $\nu$ be a binary variable where $\nu =0$ indicates if the
individual belongs to the susceptible group (the individual will
eventually experience the event of interest if followed for long
enough) and $\nu =1$ indicates if the subject is cured (the
individual will never experience the event). The proportion of cured
patients and the survival function in the group of uncured patients
can depend on certain characteristics of the subject, represented by
a set of covariates $\mathbf{X}$. Let $p(\mathbf{x}) = P(\nu =0
|\mathbf{X}=\mathbf{x})$ be the conditional probability of not being cured, and let $Y$ be the time to occurrence of the event. When $\nu=1$ it is assumed that $Y = \infty$.

The conditional distribution function of $Y$ is $F(t|\mathbf{x})=P(Y\leq t |
\mathbf{X}=\mathbf{x})$. Note that the corresponding survival function, $S(t|\mathbf{x})$, is improper when cured patients exist, since $\lim_{t \rightarrow \infty} S(t|\mathbf{x}) = 1 - p(\mathbf{x}) >0$. The conditional survival function of $Y$ given that the subject is not cured is denoted by
\begin{displaymath}
S_0(t|\mathbf{x})=P(Y > t | \mathbf{X}=\mathbf{x}, \nu =0).
\end{displaymath}
Then, the mixture cure model can be written as:
\begin{equation}  \label{cheda:supervivencia_mixtura}
S(t|\mathbf{x})=1 - p(\mathbf{x}) + p(\mathbf{x}) S_0 (t|\mathbf{x}),
\end{equation}
where $1 - p(\mathbf{x})$ is the incidence and $S_0(t|\mathbf{x})$ is the latency. We assume that each individual is subject to random right censoring and that the censoring time $C$, with distribution function $G$, is independent
of $Y$ given the covariates $\mathbf{X}$. Let $T=\min (Y,C)$ be the observed time with distribution function $H$ and $\delta = I(Y \leq C)$ the uncensoring indicator. Observe that $\delta=0$ for all the cured patients, and it also happens for uncured patients with censored lifetime. Without loss of generality, let $X$ be a univariate continuous covariate with density function $m(x)$. Therefore, the observations will be $\{(X_i, T_i, \delta_i),
i=1, \dots, n\}$ independent and identically distributed
(iid) copies of the random vector $(X, T, \delta)$.

In order to introduce the nonparametric approach in mixture cure models, we
consider the generalized Kaplan-Meier estimator by \cite{Beran} to estimate
the conditional survival function with covariates:
\begin{equation}  \label{cheda:S_est}
\hat S_{h}(t|x) = \prod_{T_{(i)} \leq t} \left ( 1 - \frac{\delta_{(i)} B_{h(i)}
(x) }{\sum_{r=i}^n B_{h(r)}(x)}\right ),
\end{equation}
where
\begin{equation}  \label{NW_weights}
B_{h(i)}(x) = \frac{K_h (x - X_{(i)})}{ \sum_{j=1}^n K_h(x - X_{(j)})}
\end{equation}
are the Nadaraya-Watson (NW) weights with $K_h(\cdot)=\frac{1}{h}K \left (
\frac{ \cdot}{h} \right )$ the rescaled kernel with bandwidth $h \rightarrow 0$. 
In the case of fixed design, the Gasser-M\"{u}ller (GM) weights \citep{Gasser} are more common. Here $T_{(1)} \leq T_{(2)} \leq \ldots \leq T_{(n)}$ are the ordered $T_i$'s, and $\delta_{(i)}$ and $X_{(i)}$ are the corresponding uncensoring
indicator and covariate concomitants. We will also denote $\hat
F_h(t|x) = 1 - \hat S_h(t|x)$ for the Beran estimator of $F(t|x)$.
The estimator (\ref{cheda:S_est}) can be extended to the case
of multiple covariates $\mathbf{X}=(X_1,\ldots,X_q)$ using, for
example, the product kernel \citep{Simonoff}. Discrete covariates can also be included by splitting the sample into subsamples corresponding to the different
category combination of the discrete covariates, for each subsample
conducting a nonparametric regression on the continuous covariates.
Another possibility is smoothing the discrete covariates with certain kernel functions \citep{Li3}.

Departing from the Beran estimator, \cite{Xu} introduced the following kernel type estimator of the incidence:
\begin{equation}  \label{ec:p_estimation}
1-\hat p_h(x)= \prod_{i=1}^{n} \left( 1 - \frac{\delta_{(i)} B_{h(i)}(x)}{%
\sum_{r=i}^n B_{h(r)}(x) } \right ) = \hat S_h(T_{\max}^1|x),
\end{equation}
where $T_{\max}^1=\max\limits_{i:\delta_i=1}(T_i)$ is the largest uncensored
failure time, and proved its consistency and asymptotic normality.

Using (\ref{cheda:supervivencia_mixtura}), we propose the following nonparametric estimator of the latency:
\begin{equation}  \label{ec:S0_estimation}
\hat S_{0,b} (t|x) = \frac{\hat S_{b}(t|x) - (1 - \hat{p}_{b}(x)}{ \hat{p}_{b}(x)},
\end{equation}
where $\hat S_{b}(t|x)$ is the Beran estimator of $S(t|x)$ in (\ref{cheda:S_est}), $1 - \hat{p}_{b}(x)$ is the estimator by \cite{Xu} in (\ref{ec:p_estimation}) and $b$ is a smoothing parameter not necessarily equal to $h$ in (\ref{ec:p_estimation}).

The identifiability of a cure model is needed to obtain unique estimates of the model functions. In a cure model, all observed uncensored lifetimes $\left( \delta _{i}=1\right)$ correspond necessarily to uncured subjects $\left( \nu_{i}=0\right)$; but it is impossible to distinguish if a subject with a censored time $\left( \delta _{i}=0\right)$ belongs to the susceptible group $\left( \nu_{i}=0\right)$ or to the non-susceptible group $\left(\nu_{i}=1\right)$, because some censored subjects may \mbox{experience} \mbox{failures} beyond the study period. This leads to difficulties in \mbox{making} a dis\-tinc\-tion between models with high incidence and long tails of the latency distribution, and low incidence and short tails of the latency distribution. To address this problem, we present Lemma \ref{lemma:identifiability}.

\begin{lem}
\label{lemma:identifiability} Let $D$ be the support of $X$. Model (\ref{cheda:supervivencia_mixtura}), with $p\left(x\right) $ and $S_{0}\left(t|x\right)$ unspecified, is identifiable if $S_{0}\left( t|x\right)$ is a proper survival function for $x\in D$.
\end{lem}

Since the proof is straightforward, it is omitted.

\subsection{Theoretical properties}
\label{subsec:2.2} 

\nolinenumbers

The Beran estimator of the conditional survival function has been
deeply studied in the literature. \cite{Dabrowska1}, in Theorem 2.1,
shows its asymptotic unbiasedness, considering NW weights. Furthermore, using GM weights, \cite{Gonzalez-Manteiga} give an almost sure iid representation for the estimator, and \cite{VanKeilegom1} prove an asymptotic repre\-sen\-ta\-tion for the bootstrapped estimator and obtain the strong consistency of the bootstrap approximation for the conditional distribution function.

Let $\widehat{\Lambda}_h(t|x)$\ be the estimator of the conditional
cumulative hazard function:
\begin{equation*}
\widehat{\Lambda}_h(t|x) = \sum_{i=1}^n \frac{\delta_{(i)} B_{h(i)}(x)}{%
\sum_{r=i}^n B_{h(r)}(x)} I(T_{(i)} \leq t) = \int_0^t \frac{d \hat
H^1_h(v|x)}{1 - \hat H_h(v^-|x)},
\end{equation*}
where
\begin{equation*}
\hat{H}_{h}(t|x)=\sum_{i=1}^{n}B_{h i}(x) I(T_{i}\leq t )
\end{equation*}
and
\begin{equation*}
\hat{H}_{h}^{1}(t|x)=\sum_{i=1}^{n}B_{h i}(x) I(T_{i}\leq t,\delta _{i}=1)
\end{equation*}%
are the empirical estimators of
\begin{equation*}
H(t|x) =P\left( T\leq t| X= x\right) \text{ \ and \ }H^{1}(t|x) =P\left(
T\leq t,\delta =1|X=x\right),
\end{equation*}
respectively. Let us define: $\tau_{H}(x) = \sup \left\{t:H(t|x) < 1\right\}$,
$\tau_{S_{0}}(x) = \sup \left\{t: \right.$ 
$ \left. S_{0}(t|x)> 0\right\}$ and $\tau_{G}(x) = \sup \left\{t:G(t|x) < 1\right\}$. Since $S( t|x)$ is an improper survival function, then $S(t|x)>0$ for any \mbox{$t \in [0, \infty)$}, and $1-H(t|x)=S(t|x)\times \bar G(t|x)$ with $\bar G(t|x) = 1 - G(t|x)$ the proper conditional survival function of the censoring time $C$, we have  $\tau_{H}(x)=\tau_{G}(x)$.

Let $\tau _{0}=\sup_{x\in D}\tau _{S_{0}}(x) $. As in \cite{Xu}, we
assume
\begin{equation}  \label{ec:tau0_tauG}
\tau _{0}<\tau _{G}\left( x\right), \forall x \in D.
\end{equation}

This condition states that the support of the censoring variable is
not contained in the support of $Y$, which guarantees that censored
subjects beyond the largest observable failure time are cured.
Hence, our estimator does not overestimate the true cure rate. A
si\-mi\-lar assumption was used by \cite{Maller1,Maller2} in
homogeneous cases. As pointed out in \cite{Laska}, if the censoring
variable takes va\-lues always below a time $\tau_G < \tau_0$, for
example in a clinical trial with a fixed maxi\-mum follow-up period,
the largest uncensored observation $T^1_{\max}$ may
occur at a time not larger than $\tau_G$ and therefore always before $\tau_0$%
. In such a case, for a large sample size, the estimator in (\ref%
{ec:p_estimation}) is an estimator of $1-p(x) + p(x)S_0(\tau_G)$ which is
strictly larger than $1-p(x)$. This comment shows the need of con\-si\-de\-ring
the length of follow-up in the design of a clinical trial carefully, so that
$S_0(\tau_G)$ is sufficiently small to take the estimator (\ref%
{ec:p_estimation}) of $1-p(x) + p(x)S_0(\tau_G)$ as a good estimator of $%
1-p(x)$ for practical purposes. The simulations in \cite{Xu} show
that if the censoring distribution $G(t|x)$ has a heavier tail than $%
S_{0}(t|x) $, the estimates from the proposed method will tend to have
smaller biases regardless of the value of $\tau_{S_{0}}(x)$.

\cite{Maller1} dealt with the problem of testing a similar condition to (\ref{ec:tau0_tauG}) in an unconditional setting. They proposed to test $H_0: \tau_0 > \tau_G$ versus the alternative $H_1: \tau_0 \leq \tau_G$. One of the weak points of this approach is to include condition (\ref{ec:tau0_tauG}) in the alternative hypothesis. Since this is a neutral assumption, it seems more reasonable to keep (\ref{ec:tau0_tauG}) if there are no strong evidences against it. In that sense, it is more natural to include (\ref{ec:tau0_tauG}) in the null hypothesis. Apart from that, the ideas by \cite{Maller1} can be extended to a conditional setting as follows. Let us consider $\Pi(t) = E(\delta|T=t)$ and define $\underline{\tau}_G = \inf_{x \in D} \tau_G(x)$. Condition (\ref{ec:tau0_tauG}) implies that $\exists a < \underline{\tau}_G$ such that $\Pi(t)=0\; \forall t \geq a$. Consequently, this condition can be checked in practice by the following hypothesis test:
\begin{equation*}
  \left\lbrace
  \begin{array}{l}
     H_0: \exists a < \underline{\tau}_G \;/\; \Pi(t)=0, \; \forall t \geq a \\
     H_1: \forall a < \underline{\tau}_G, \; \exists t \geq a \;/\; \Pi(t) > 0
  \end{array} .
  \right.
\end{equation*}
This can be tested by means of nonparametric regression estimators of $\Pi(t)$ based on the sample $(T_1, \delta_1), \dots, (T_n, \delta_n)$. To do that, it is necessary to estimate $\underline{\tau}_G$ in a nonparametric way. This can be done by just estimating the support of every $G(t|x)$, via the Beran estimator.

We need to consider the following assumptions, to be used in the asymptotic results for the incidence estimator:
\begin{enumerate}[({A}1)]
\item $X$, $Y$ and $C$ are absolutely continuous random variables.
\item Condition (\ref{ec:tau0_tauG}) holds.
\item \begin{enumerate}[(a)]
\item Let $I=[x_{1},x_{2}]$ be an interval contained in the support of $%
m$, and $I_{\delta }=[x_{1}-\delta ,x_{2}+\delta ]$ for some $\delta >0$
such that
\begin{equation*}
0<\gamma =\inf [m\left( x\right) :x\in I_{\delta }]<\sup [m\left( x\right)
:x\in I_{\delta }]=\Gamma <\infty
\end{equation*}%
and $0<\delta \Gamma <1$. For all $x\in I_{\delta}$ the random variables
$Y$ and $C$ are con\-di\-tio\-na\-lly independent given $X=x$.
\item There exist $a,b\in \mathbb{R}$, with $a<b$ satisfying $1 - H(t|x) \geq \theta > 0$ for $(t,x) \in [a,b] \times I_{\delta}$.
\end{enumerate}
\item The first derivative of the function $m(x)$ exists and is
continuous in $x\in I_{\delta }$ and the first derivatives with respect to $x
$ of the functions $H(t|x)$ and $H^{1}(t|x)$ exist and are continuous and
bounded in $\left(t,x\right) \in \lbrack 0,\infty )\times I_{\delta }$.
\item The second derivative of the function $m(x)$ exists and is %
\mbox{continuous} in $x\in I_{\delta }$ and the second derivatives with
respect to $x$ of the functions $H(t|x) $ and $H^{1}(t|x) $ exist and are
continuous and bounded in $\left(t,x\right) \in \lbrack 0,\infty )\times
I_{\delta }$.
\item The first derivatives with respect to $t$ of the functions $%
G(t|x)$, $H(t|x)$, $H^{1}(t|x)$ and $S_0(t|x)$ exist and are continuous in $%
\left( t,x\right) \in \lbrack a,b] \times D$.
\item The second derivatives with respect to $t$ of the functions $%
H(t|x) $ and $H^{1}(t|x) $ exist and are continuous in $\left(t,x\right) \in
\lbrack a,b]\times D$.
\item The second partial derivatives with respect to $t$ and $x$ of the functions $H(t|x)$ and $H^1(t|x)$ exist and are continuous and bounded for $(t,x)\in [0,\infty) \times D$.
\item Let us define $H_{c,1}(t)=P(T < t |\delta=1)$. The first and second derivatives of the distribution and subdistribution functions $H(t)$ and $H_{c,1}(t)$ are bounded away from zero in $[a,b]$. Moreover, $H^{\prime }_{c,1}(\tau_0) > 0$.
\item The functions $H(t|x)$, $S_{0}(t|x)$ and $G(t|x)$ have bounded second-order derivatives with respect to $x$ for any given value of $t$.
\item $\mathlarger \int_0^{\infty} \dfrac{dH^1(t|x)}{(1-H(t|x))^2} < \infty$ $\forall x \in I$.
\item The kernel function, $K$, is a symmetric density vanishing outside
$\left(-1,1\right) $ and the total variation of $K$ is less than some $%
\lambda <$ $\infty $.
\item The density function of $T$, $f_T$, is bounded away from 0 in $[0,\infty)$.
\end{enumerate}

Assumptions $\left( A1\right), \left( A3\right) -\left( A9  \right)$ and $(A12)-(A13)$ are necessary in Theorem \ref{thm:iid} because its proof is strongly based on Theorem 2 of \cite{Iglesias-Perez}. Assumptions (A2) and (A10) are needed to prove Lemma \ref{Th_Tn_conv_tau0} and, consequently, required so existing results in the literature, stated for a fixed $t$ such that $1 - H(t|x) \geq \theta > 0$ in $(t,x) \in [a,b] \times I_{\delta}$, can be applied with the random value $t= T_{\max }^{1}$. Assumption (A11) is necessary to bound the result of an integral in Lemma \ref{TH_iidL}.

In the next theorem we show that both the proposed nonparametric incidence and latency estimators are the local maximum likelihood estimators of $1 -p(x)$
and $S_0(t|x)$.

\begin{thm}
\label{thm:mle} The estimators $1 - \hat{p}_h\left(x\right) $ and $\hat{S}_{0,b}(t|x),$ given in (\ref{ec:p_estimation}) and (\ref{ec:S0_estimation}) respectively, are the local maximum likelihood estimators of $1 - p\left(x\right) $ and $S_{0}(t|x)$ for the mixture cure model (\ref{cheda:supervivencia_mixtura}), for any $x \in D$ and $t \geq 0$.
\end{thm}

We also obtain an iid representation of the incidence estimator.

\begin{thm}
\label{thm:iid} Under assumptions $\left( A1\right) - \left( A13\right) $, for any sequence of bandwidths satisfying $nh^{5}(\ln n)^{-1}=O(1)$ and $\ln n / (nh) \rightarrow 0$, then
\begin{equation*}
(1-\hat{p}_{h}(x))-(1-p(x))=\left( 1-p\left( x\right) \right) \sum_{i=1}^{n}%
\tilde{B}_{hi}(x)\xi \left( T_{i},\delta _{i},x\right) +R_{n}\left( x\right),
\end{equation*}
where
\begin{equation}
\tilde{B}_{hi}(x)=\frac{\frac{1}{nh}K\left( \frac{x-X_{i}}{h}\right) }{m(x)},
\label{Btilde}
\end{equation}
\begin{equation}
\xi \left( T_{i},\delta _{i},x\right) =\frac{I(\delta _{i}=1)}{1-H(T_{i}|x)}%
-\int_{0}^{T_{i}}\frac{dH^{1}(t|x)}{\left( 1-H(t|x)\right) ^{2}}  \label{xi}
\end{equation}%
and
\begin{equation*}
\sup_{x\in I}|R_{n}(x)|=O\left( \left( \frac{\ln n}{nh}\right) ^{3/4}\right)
\text{ a.s.}
\end{equation*}
\end{thm}

Finally, from the representation in Theorem \ref{thm:iid} with
straightforward calculations, the asymptotic expression of the mean
squared error of the incidence estimator,
\begin{equation}  \label{ec:MSE}
MSE_x(h_x)=E[(\hat{p}_{h_x}(x)-p(x))^2],
\end{equation}
is given by:
\begin{equation}  \label{eq:AMSE}
AMSE_x(h_{x}) = \frac{1}{nh_{x}} (1 - p(x))^2 c_K \sigma^2(x) + \left [ h_{x}^2 \frac{1}{2} d_K (1-p(x)) \mu(x) \right ]^2,
\end{equation}
where the first term corresponds to the asymptotic variance and the second
one to the asymptotic squared bias, with $d_K = \int v^2 K(v) dv$, %
\mbox{$c_K = \int K^2(v) dv$} and, following a notation similar to that in
\cite{Dabrowska2}:
\begin{equation*}
\mu(x) = \frac{2 \Phi^{\prime }(x,x)m^{\prime }(x)+\Phi^{\prime \prime
}(x,x)m(x)}{m(x)},
\end{equation*}
\begin{equation*}
\sigma^2(x) = \frac{1}{m(x)} \int_0^{\infty} \frac{dH^1(t|x)}{(1 - H(t|x))^2}%
,
\end{equation*}
where
\begin{equation*}
\Phi(u,x) = \int_0^{\infty} \frac{dH^1(t|u)}{1 - H(t|x)} - \int_0^{\infty}
\frac{1 - H(t|u)}{(1 - H(t|x))^2} dH^1(t|x),
\end{equation*}
with $\Phi^{\prime}(u,x) = \partial / (\partial u) \Phi(u,x)$ and $%
\Phi^{\prime\prime}(u,x) = \partial^2 / (\partial u^2) \Phi(u,x)$.

\section{Bandwidth selection}
\label{sec:3} 

\nolinenumbers

The choice of the bandwidth is a crucial issue in kernel estimation, since it controls the trade-off between bias and variance. 
Most of the methods for smoothing parameter selection in nonparametric curve estimation look for a small error when approximating the underlying curve by the smooth estimate. The
asymptotically optimal local bandwidth to estimate the cure rate, $1-p(x)$,
in the sense of minimizing the asymptotic expression of the $MSE_x$ in (\ref%
{eq:AMSE}), is given by:
\begin{equation*}
h_{x,AMSE} = \left ( \frac{c_K \sigma^2(x)}{d_K^2 \mu^2(x)} \right )^{1/5}
n^{-1/5},
\end{equation*}
which is an asymptotic approximation of the bandwidth $h_{x,MSE}$ that minimizes the $MSE_x$. The optimal bandwidth $h_{x,AMSE}$ depends on unknown functions through $\mu(x)$ and $\sigma^2(x)$. Considering \cite{Dabrowska1}, a plug-in bandwidth selector can be obtained by replacing those unknown functions by consistent nonparametric estimates, 
giving rise to a never-ending process, which seems even
harder than the original problem of incidence estimation. On the other hand,
unfortunately, the finite-sample behavior of the cross validation (CV)
bandwidth selector in this context turned out to be disappointing. The CV
bandwidth was highly variable and tended to undersmooth (results not shown).

\subsection{Bootstrap bandwidth selector}
\label{sec:3_1} 

\nolinenumbers

Another way to select the bandwidth is to use the bootstrap method. It consists of mi\-ni\-mi\-zing a bootstrap estimate of the mean squared error, $MSE_x(h_x)$.

We consider the simple weighted bootstrap, without resampling the covariate $X$, which is equivalent to the simple weighted bootstrap proposed by \cite{Li2}. For fixed $x$ and $i =1,\ldots, n$, we set $X_i^* = X_i$ and generate a pair $( T_i^*, \delta_i^*)$ from the weighted empirical distribution $\hat F_{g_x}(\cdot,\cdot | X_i^*)$, where
\begin{equation*}
\hat F_{g_x}(u, v | x) = \sum_{i=1}^n B_{g_x i}(x) I (T_{i}\leq u,
\delta_{i} \leq v)
\end{equation*}
and $B_{g_x i}(x)$ is the NW weight in (\ref{NW_weights}) with pilot
bandwidth $g_x$. The resulting bootstrap resample is $\{ (X_1, T_1^*,
\delta_1^*), \dots, (X_n, T_n^*, \delta_n^*)\}$. From now on, we will use the notation $E^*$ and $P^*$ for bootstrap expectation and probability, i.e., conditionally on the original observations.

The bootstrap bandwidth is the minimizer of the bootstrap version of \\ $MSE_x(h_x)$ in (\ref{ec:MSE}),
\begin{equation}  \label{eq:MSEb1}
MSE_{x,g_x}^*(h_x) = E^* [(\hat p_{h_x, g_x}^*(x) - \hat p_{g_x}(x) )^2 ],
\end{equation}
which consists of replacing the original sample by the bootstrap resample, the kernel incidence estimator based on the sample by its bootstrap version and the theoretical incidence function by the estimated incidence based on a pilot bandwidth, $g$. It can be approximated, using Monte Carlo, by:
\begin{equation}  \label{eq:MSEb}
MSE_{x,g_x}^*(h_x) \simeq \frac{1}{B} \sum_{b=1}^B (\hat p_{h_x, g_x}^{*b}(x) - \hat p_{g_x}(x) )^2,
\end{equation}
where $\hat p_{h_x, g_x}^{*b}(x)$ is the kernel estimator of $p(x)$ using
bandwidth $h_x$ and based on the $b$-th bootstrap resample generated from $%
\hat F_{g_x}$, and $\hat{p}_{g_x}(x)$ is the kernel estimator of $p(x)$
computed with the original sample and pilot bandwidth $g_x$.

Considering a bandwidth search grid $\{h_1,\ldots,h_L\}$, the procedure for obtaining the bootstrap bandwidth selector for a fixed covariate value, $x$, is as follows:

\begin{enumerate}[1.]
\item Generate $B$ bootstrap resamples of the form: \\ $\big \{ \left (
X_1^{(b)}, T_1^{*(b)}, \delta_1^{*(b)} \right ), \dots, \left ( X_n^{(b)},
T_n^{*(b)}, \delta_n^{*(b)} \right ) \big \}$, $b=1, \dots, B$.
\item For the $b$-th bootstrap resample $(b = 1, ... ,B)$, compute the
nonparametric estimator $\hat{p}_{h_l,g_x}^{\ast b}(x)$ with bandwidth $h_l$, $%
l = 1, 2, \ldots , L$.
\item With the original sample and the pilot bandwidth $g_x$, compute $\hat
p_{g_x}(x)$.
\item For each bandwidth $h_l$ in the grid, compute the Monte Carlo
approximation of $MSE_{x,g_x}^*(h_l)$, given by (\ref{eq:MSEb}).
\item The bootstrap bandwidth, $h_x^*$, is the minimizer of the Monte
Carlo approximation of $MSE_{x,g_x}^*(h_l)$ over the grid of bandwidths $%
\{h_1,\ldots,h_L\}$.
\end{enumerate}

Based on the results in \cite{VanKeilegom1,VanKeilegom2} for fixed design with GM weights, the optimal pilot bandwidth, $g_x$, could be chosen so that it minimizes (\ref{eq:MSEb1}) for a given sample. However, simulation results showed that the choice of the pilot bandwidth has a small effect on the final bootstrap bandwidth. Consequently, a simple rule is proposed to select $g_x$ (see equation (\ref{eq:g_dat_sim}) below).

\paragraph{Remark} The bandwidth sequence $g_x=g_{n}$ has to be typically
asymptotically larger than $h_x=h_{n}$. This oversmoothing pilot bandwidth
is required for the bootstrap bias and variance to be asymptotically
efficient estimators for the bias and variance terms. The order $n^{-1/9}$ for this asymptotically optimal pilot bandwidth satisfies the conditions in Theorem 1 of \cite{Li2}, and it coincides with the order obtained by \cite{Cao2} for the uncensored case.

\section{Simulation study}
\label{sec:4} 

\nolinenumbers

In this section we compare the proposed nonparametric estimators with
the semi\-para\-metric estimators in \cite{Peng1}, which are implemented
in the \emph{smcure} package in R \citep{Cai}. These estimators assume
a logistic expression for the incidence and a proportional hazards (PH)
model for the latency.

We carry out a simulation study with two purposes. First, we evaluate the
finite sample performance of the nonparametric estimators $1-\hat{p}_{h_x}$
and $\hat S_{0,b_x}$, both computed in a grid of bandwidths with the Epanechnikov kernel, and we compare the results with those of
the semiparametric estimators. Second, the practical behavior of the
bootstrap bandwidth selector is assessed. We consider two different models
and for both, the censoring times are generated according
to the exponential distribution with mean $1/0.3$ and the covariate $X$ is $U(-20,20)$.

\begin{figure}[tbp]
\includegraphics[width=0.5\textwidth]{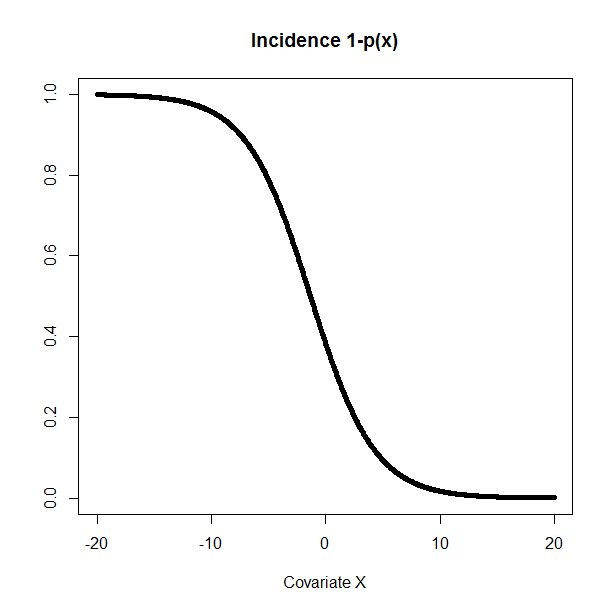} %
\includegraphics[width=0.5\textwidth]{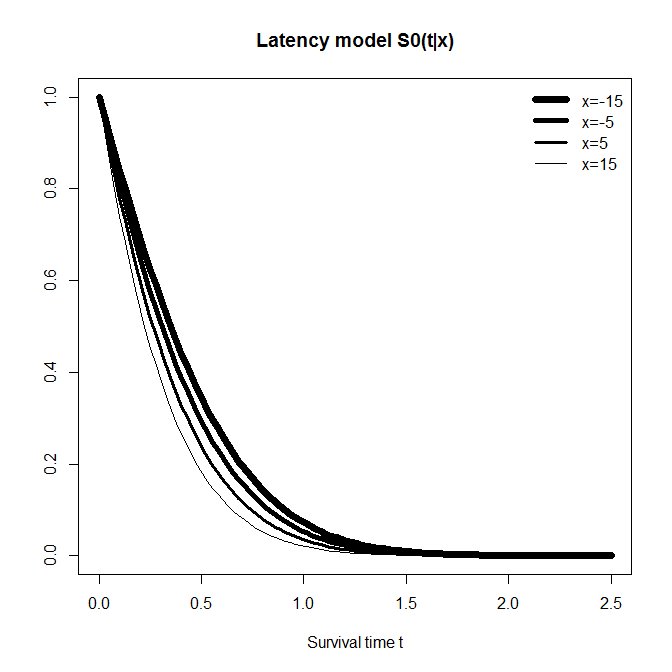} \newline
\includegraphics[width=0.5\textwidth]{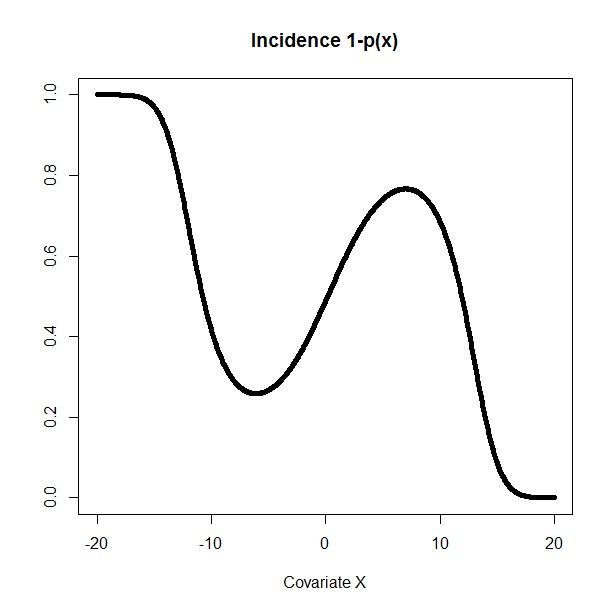} %
\includegraphics[width=0.5\textwidth]{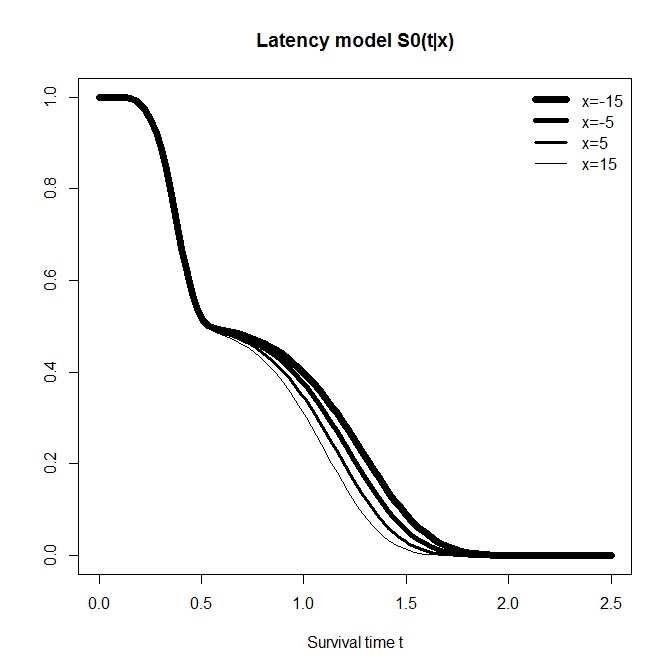}
\caption{Theoretical incidence (left) and latency (right) functions in Model 1 (top) and Model 2 (bottom).}
\label{p_latencia_teorica_M2}
\end{figure}

\paragraph{Model 1} For comparison reasons, this simulated setup is
the same as the so-called mixture cure (MC) model considered in \cite{Xu}. The data are generated from a logistic-exponential MC model, where
the probability of not being cured is
\begin{equation*}
p(x) = \frac{ \exp ( \beta_0 + \beta_1 x ) }{ 1 + \exp (\beta_0 + \beta_1 x
) },
\end{equation*}
with $\beta_0 = 0.476$ and $\beta_1 = 0.358$, and the survival function of
the uncured subjects is:
\begin{equation*}
S_0(t|x) =
\begin{cases}
\dfrac{ \exp(-\lambda(x) t) - \exp(-\lambda(x) \tau_0) }{1 -
\exp(-\lambda(x) \tau_0)} & \mbox{if } t \leq \tau_0 \\
0 & \mbox{if } t > \tau_0%
\end{cases},
\end{equation*}
where $\tau_0 = 4.605$ and $\lambda \left( x\right) =\exp \left( (x+20) /40
\right) $. The percentage of censored data is 54\% and of
cured data is 47\%. In Figure \ref{p_latencia_teorica_M2} (top) we
show the shape of the theoretical incidence and latency
functions. Note that in this model the
incidence is a logistic function and the latency is a function which
is very close to fulfill the proportional hazards model and that has been truncated
to guarantee condition (\ref{ec:tau0_tauG}). Therefore, the semiparametric estimators are expected to give very good results in this model.

\paragraph{Model 2} The data are generated from a cubic logistic-exponential
mixture model, where the incidence is:
\begin{equation*}
1 - p(x) = 1 -  \frac{\exp \left ( \beta_0 + \beta_1 x + \beta_2 x^2 + \beta_3 x^3
\right  ) }{ 1 + \exp \left (\beta_0 + \beta_1 x + \beta_2 x^2 + \beta_3 x^3
\right) },
\end{equation*}
with $\beta_0 = 0.0476$, $\beta_1 = - 0.2558 $, $\beta_2 = - 0.0027$ and $%
\beta_3 = 0.0020$, and the latency is:
\begin{equation*}
S_0 (t|x) = \frac{1}{2} \left ( \exp(-\alpha(x) t^5 ) + \exp(-100 t^5)
\right ),
\end{equation*}
with
\begin{equation*}
\alpha(x) = \frac{1}{5} \exp((x+20)/40).
\end{equation*}
The percentages of censored and cured data are 62\% and 53\%,
respectively. Figure \ref{p_latencia_teorica_M2} (bottom) shows the theoretical incidence and latency in this model. The incidence is not a logistic function and the effect of the covariate on the failure time of the uncured patients does not fit a PH model anymore. So, the results will show the gain of using
the proposed nonparametric estimators, that do not require any parametric or semiparametric assumptions, with respect to the semiparametric ones.

\subsection{Efficiency of the nonparametric estimators}
\label{subsec:4_1}

\nolinenumbers

\begin{figure}[tbp]
\includegraphics[width=0.5\textwidth]{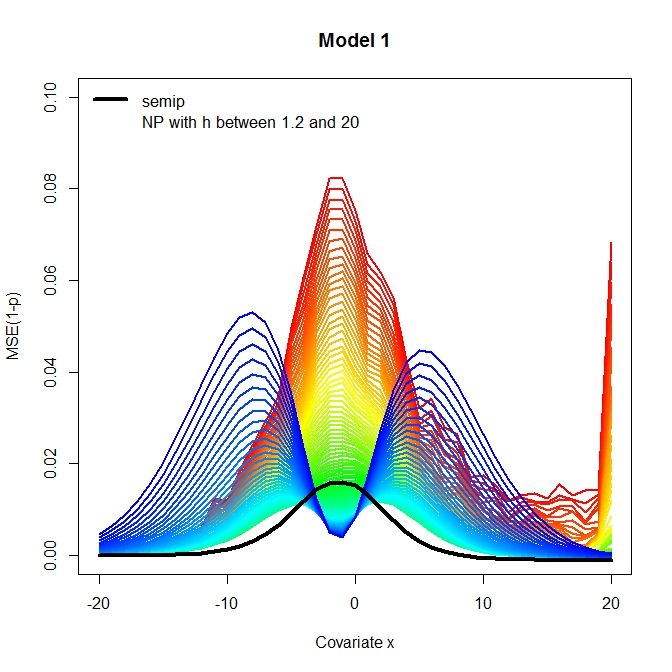} %
\includegraphics[width=0.5\textwidth]{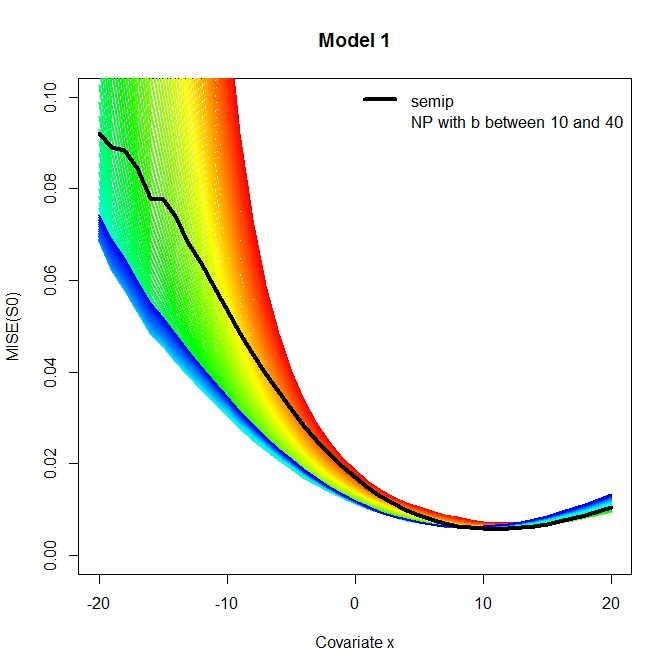} \newline
\includegraphics[width=0.5\textwidth]{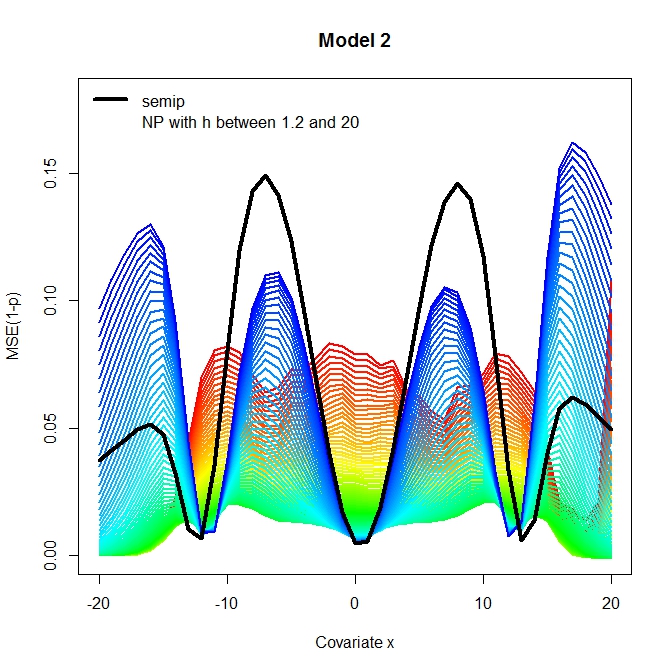} %
\includegraphics[width=0.5\textwidth]{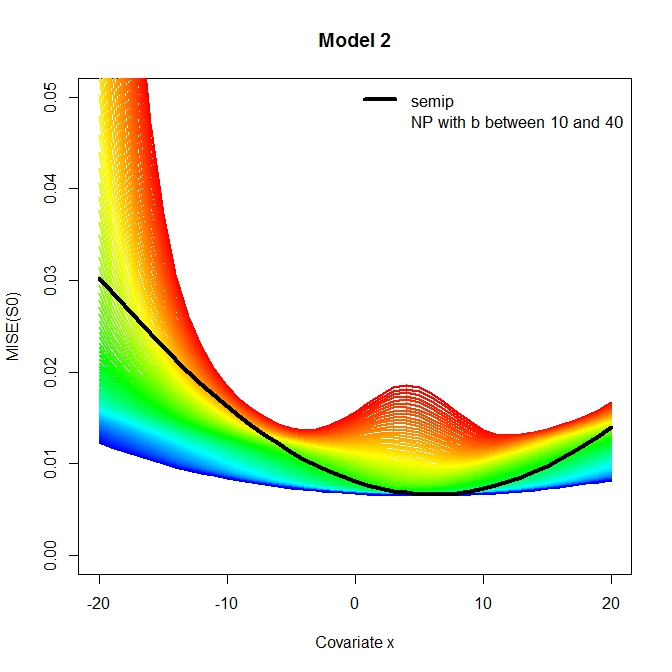}
\caption{On the left, MSE for the semiparametric (black line) and
the nonparametric estimators of $1 - p(x)$ computed with different
bandwidths: from $h_1=1.2$ (red line) to $h_{100}=20$ (dark
blue line). On the right, MISE for the semiparametric (black line)
and the nonparametric estimators of $S_0(t|x)$ computed with
different bandwidths: from $b_1=10$ (red line) to $b_{100}=40$
(dark blue line). The data were generated from Model 1 (top)
and Model 2 (bottom).} \label{MSE_MISE_M1}
\end{figure}

A total of $m=1000$ samples of size $n=100$ are drawn to approximate, by Monte Carlo, the mean squared error (MSE) of the incidence estimators, and the mean integrated squared error (MISE) of the latency estimators, for a grid of 100 bandwidths in a logarithm scale, from $h_1=1.2$ to $h_{100}=20$ for the incidence function, and from $b_1=10$ to $b_{100}=40$ for the latency. The results for both models are shown in Figure \ref{MSE_MISE_M1}.

Regarding the MSE of the incidence estimators, Figure \ref{MSE_MISE_M1} shows that in Model 1 there is a range of bandwidths, from $h_{50}=4.83$ to $h_{70}=8.53$ (light blue lines) for which the nonparametric estimator is quite competitive with respect to the semiparametric estimator in values $x$ of the covariate near the endpoints of the support of $X$, and it works much better when the value of the covariate is around $0$. In Model 2, as expected, the
nonparametric estimator outperforms the semiparametric one for a wide range of bandwidths except for $3$ singular values of the covariate $x$, those for which $p^{\prime \prime}(x)=0$ and the semiparametric estimate is very close to the true value of $(1-p(x))$. These $3$ values are specific points in which the semiparametric estimation passes through the theoretical function.

Considering the latency estimators, it is noteworthy that in Model 1, for values of the covariate from $x=-20$ to $x=10$, there is also a very wide range of bandwidths, specifically, between $b_{30}=15.01$ (light green lines) and $b_{100}=40$ (dark blue lines), for which the MISE of the nonparametric estimator is smaller than the MISE of the semiparametric estimator, it can be seen in Figure \ref{MSE_MISE_M1} (top, right).
  In Model 2, the nonparametric estimator of the latency, computed with bandwidths between $b_{20}=13.05$ (yellow lines) to $b_{100}=40$ (dark blue lines), outperforms the semi\-para\-metric estimator for all the covariate values, except for $x \in [4,9]$, where the semiparametric estimator is very competitive. 
In short, the nonparametric estimators are quite comparable to the semiparametric ones in situations where the latter are expected to give better results, as in Model 1, and they outperform the semiparametric estimators when the incidence is not a logistic function and the latency does not fit a PH model (Model 2).
The efficiency of the nonparametric estimators depends on the choice of the bandwidth, but although the optimal value of the bandwidth remains unknown, the simulations show that, for quite wide ranges of bandwidths, the proposed nonparametric methods outperform the existing semiparametric estimator by \cite{Peng1}.

\subsection{Efficiency of the bootstrap bandwidth selector for the incidence}
\label{subsec:4_2}

\nolinenumbers

In this simulation study, we consider sample sizes of $n=50, 100$
and $200$. For $m=1000$ trials, we approximate the
$MSE_x$ and the optimal bandwidth, $h_{x,MSE}$, of the proposed nonparametric estimator of the incidence. The $MSE_{x,g_x}(h^*)$ and the bootstrap bandwidth $h^*_{x,g_x}$ are also approximated.

Note that minimizing $MSE_{x,g_x}^*(h_x)$ in $h_x$ for each value, $x$, of the cova\-riate, is a computationally expensive algorithm. For that reason, we carry out a two-step method with a double search in each stage. In the first step, we draw $B=80$ bootstrap resamples and consider a number of $21$ bandwidths equispaced on a logarithmic scale, from $h_1=0.2$ to $h_{21}=50$ in the first search, whereas in the second search the grid is centered around the optimal bandwidth obtained in the first search. Then, we carry out the second step with also a double search in a similar way we did for the first step, but now with two differences: we draw $B=1000$ bootstrap resamples and we consider a finer smaller grid of $5$ bandwidths in both the first and second search.

In view of the fact that the choice of $g_x$ has a
low effect on the final boots\-trap bandwidth, we
propose to use a naive selector, keeping the $n^{-1/9}$ optimal
order. Since the distribution of the covariate is uniform, we
consider the follo\-wing global pilot bandwidth, that does not
depend on the value $x$ for which the estimation is to be carried
out:
\begin{equation}  \label{eq:g_dat_sim}
g = \frac{X_{(n)} - X_{(1)}}{10^{7/9}}n^{-1/9}.
\end{equation}
Note that, for $X\in U(-20,20)$, when $n=100$ the value of the global pilot bandwidth $g$ is $(X_{(n)} - X_{(1)})/10 \simeq 4$. Similarly, $g \simeq 4.32$ ($g \simeq 3.70$) when $n=50$ ($n=200$). For a naive pilot bandwidth selector if the distribution for $X$ can not be assumed uniform, see Section \ref{sec:5}.

Figure \ref{bootstrap_MSE} shows the $MSE_x$ evaluated at the median, $25$th and $75$th percentiles of the proposed bootstrap bandwidth, along the $m = 1000$ simulated samples. The value of the $MSE_x$ for the nonparametric estimator, approximated by Monte Carlo and evaluated at the MSE bandwidth, $h_{x,MSE}$, is also given as reference. We observe that the median, $25$th and $75$th percentiles of the bootstrap bandwidths have an MSE close to the optimal value. As expected, the similarity increases with the sample size. Moreover, we can also check how $MSE_x(h_{x,MSE})$ and $MSE_{x}(h^*)$ decrease as $n$ becomes larger.

\afterpage{ 
\clearpage 
\begin{figure}[tbp]
\includegraphics[width=0.5\textwidth]{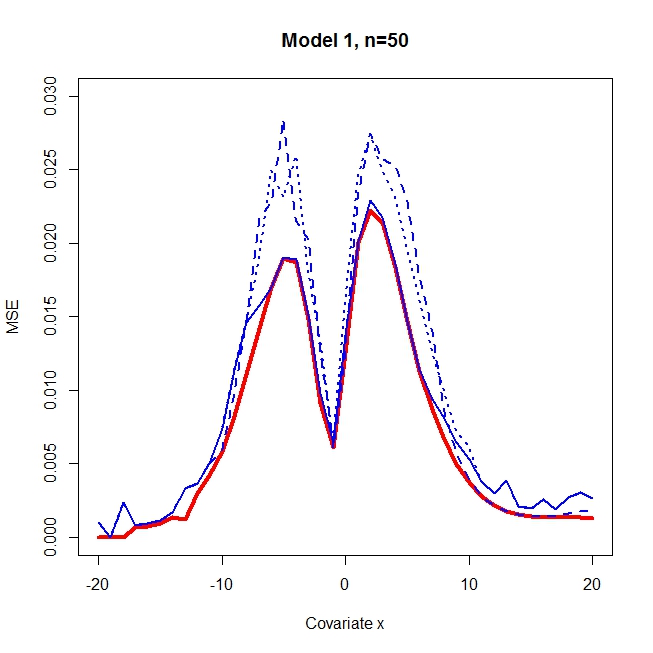} %
\includegraphics[width=0.5\textwidth]{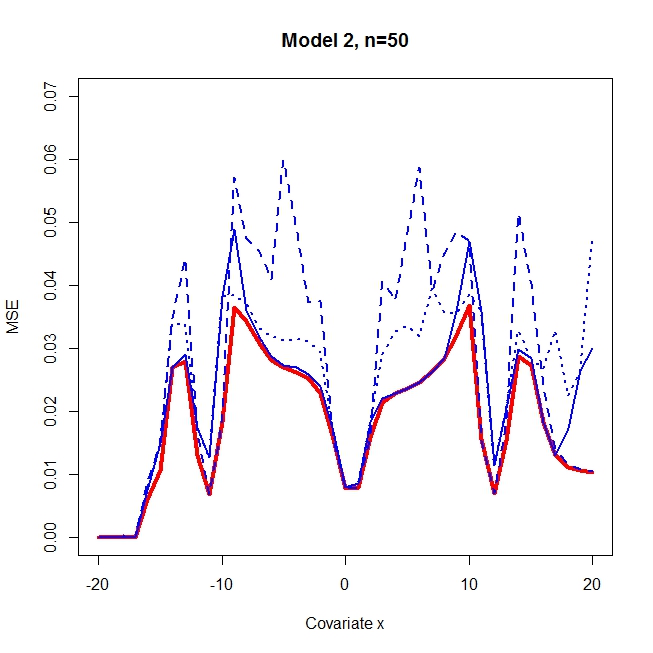} \newline
\includegraphics[width=0.5\textwidth]{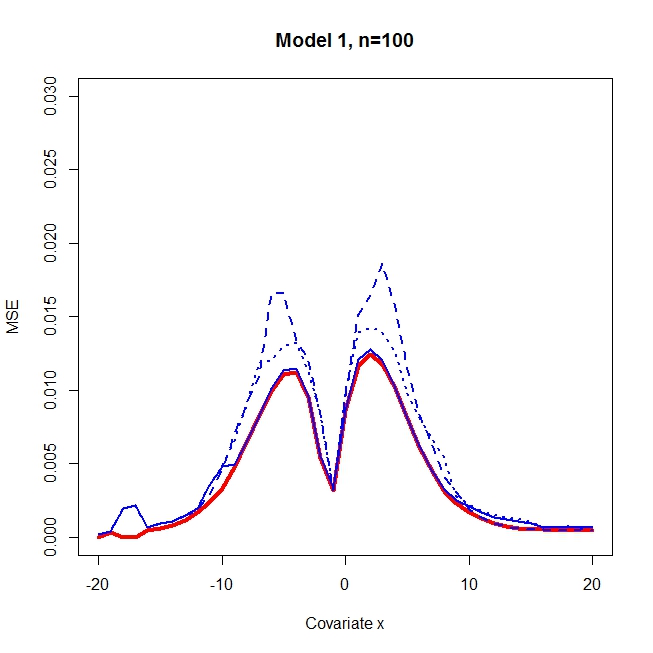} %
\includegraphics[width=0.5\textwidth]{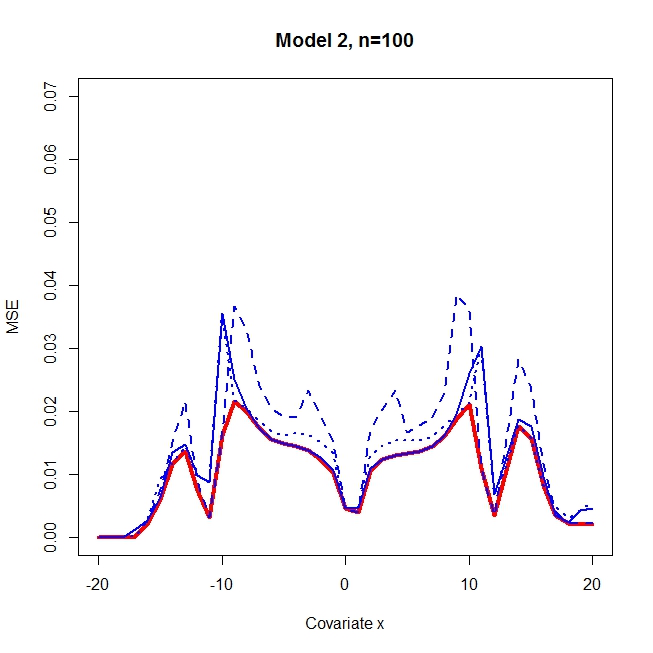} \newline
\includegraphics[width=0.5\textwidth]{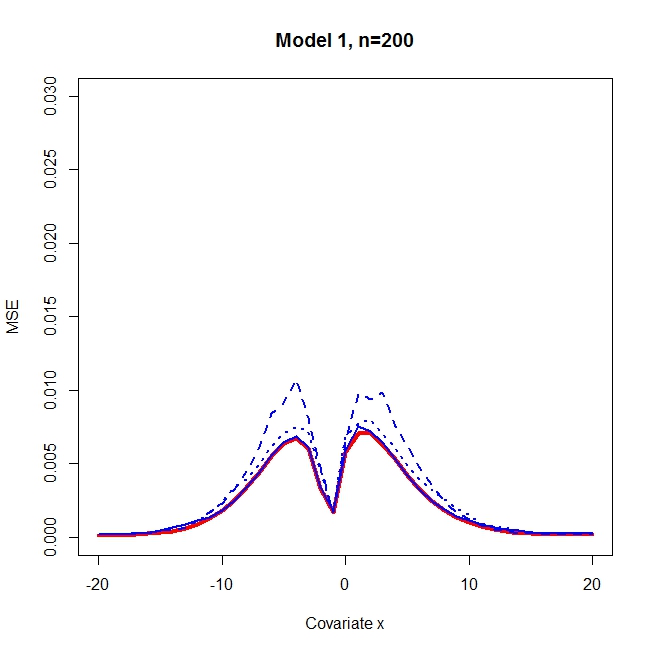} %
\includegraphics[width=0.5\textwidth]{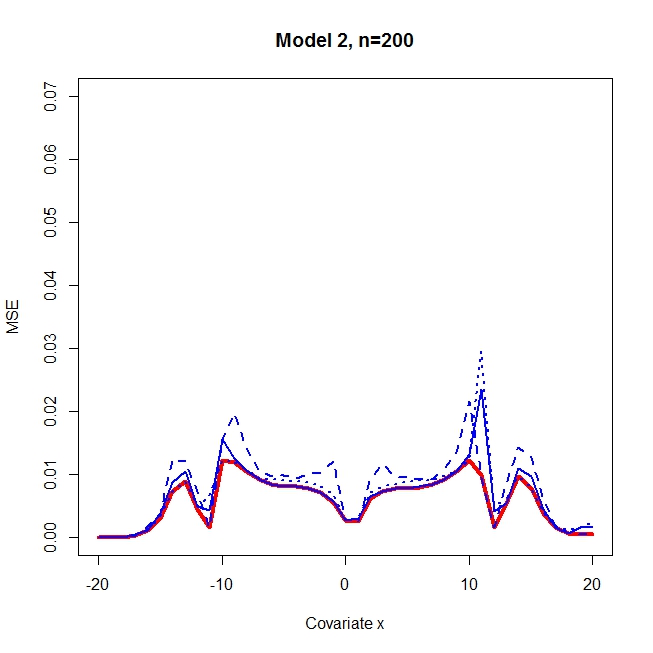} \newline
\caption{$MSE_x$ of the nonparametric estimator of the incidence evaluated at $h_{x,MSE}$ (red line), and $MSE_{x}$ evaluated at the median (solid blue line), 25th (dotted blue line) and 75th (dashed blue line) percentiles of the bootstrap bandwidth, $h_x^*$, along $m = 1000$ samples of sizes $n = 50$ (top), $n = 100$ (center) and $n = 200$ (bottom), for Model 1 (left) and Model 2 (right). } 
\label{bootstrap_MSE}
\end{figure}
\thispagestyle{empty}
\clearpage 
}

The performance of the bootstrap bandwidth for Models 1 and 2 is shown in Figure \ref{bootstrap_h}. The optimal $h_{x,MSE}$, approximated by Monte Carlo, is displayed together with the median and the 25th and 75th percentiles of the $1000$ bootstrap bandwidths, $h_x^*$. We can appreciate how the bootstrap bandwidth, $h_x^*$, approaches $h_{x,MSE}$, adapting properly to the shape of $h_{x,MSE}$ for the three sample sizes. The optimal bandwidth, $h_{x,MSE}$, has got peaks at the values $x$ of the covariate for which $p^{\prime \prime}(x)=0$. In other terms, those peaks only occur at points $x$ for which the asymptotically optimal bandwidth is infinitely large because the best choice is to smooth as much as possible, and the best local fit is a global fit. Note that if such large bandwidths are used, those values of $x$ correspond to the values where the $MSE_x$ shows deep valleys, that is, there is a noticeable improvement in the estimation of the incidence.

\afterpage{ 
\clearpage 
\begin{figure}[tbp]
\includegraphics[width=0.5\textwidth]{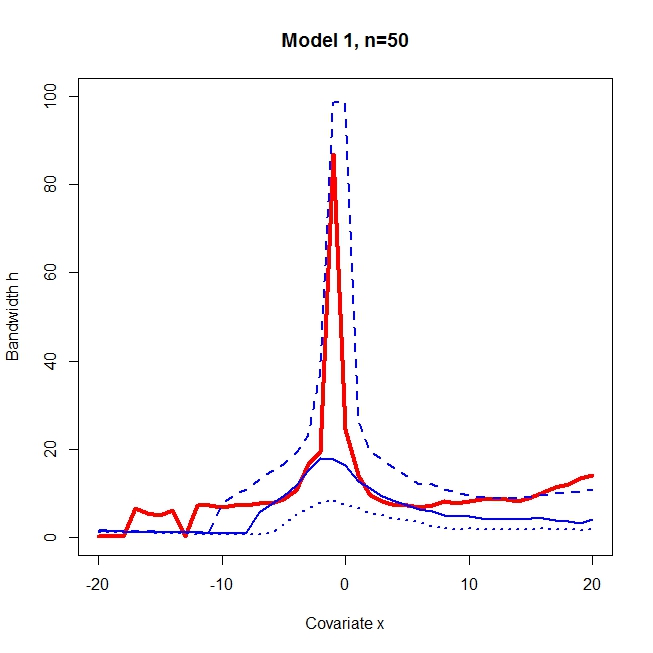} %
\includegraphics[width=0.5\textwidth]{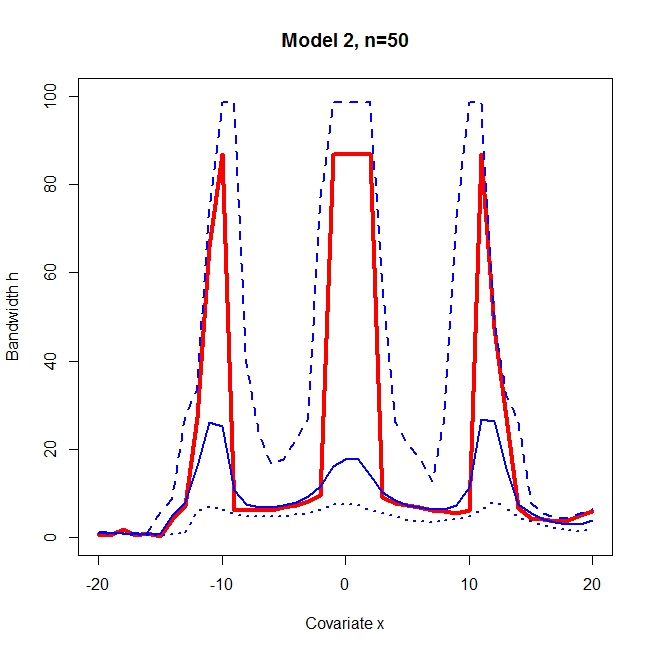} \newline
\includegraphics[width=0.5\textwidth]{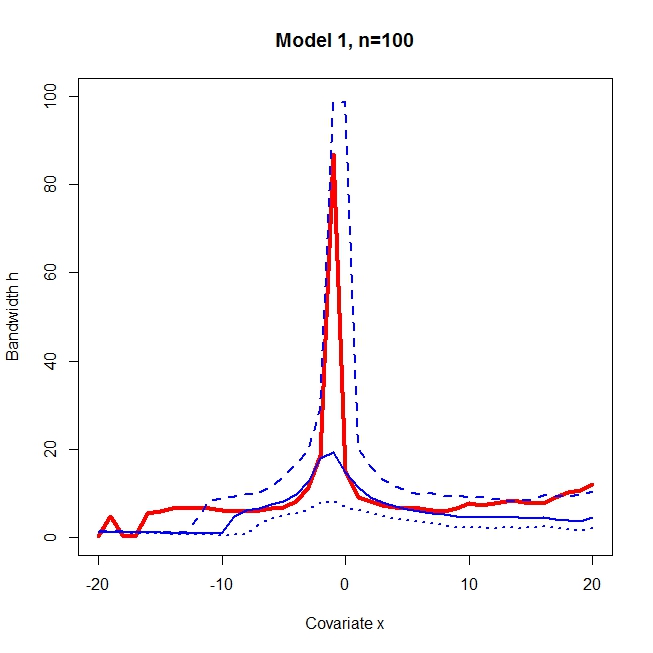} %
\includegraphics[width=0.5\textwidth]{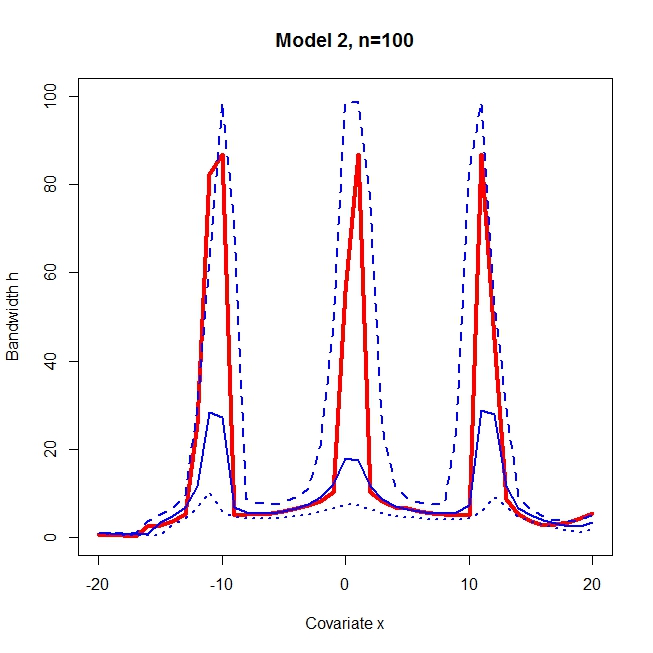} \newline
\includegraphics[width=0.5\textwidth]{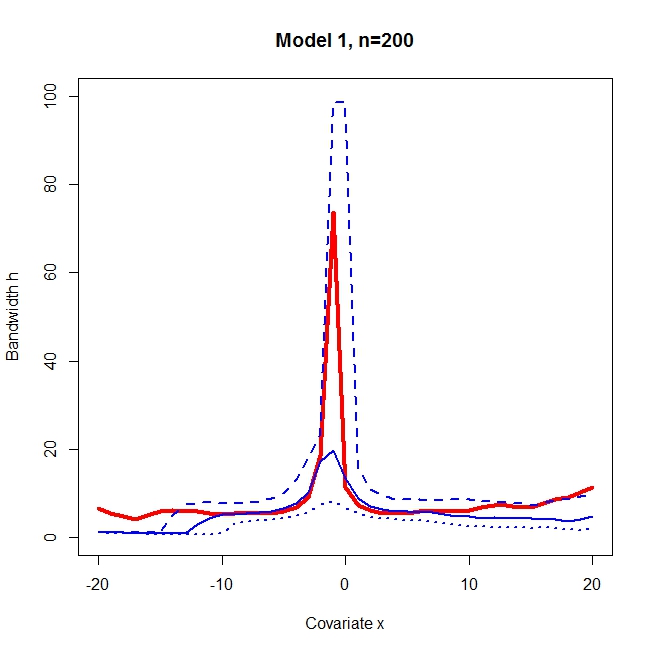} %
\includegraphics[width=0.5\textwidth]{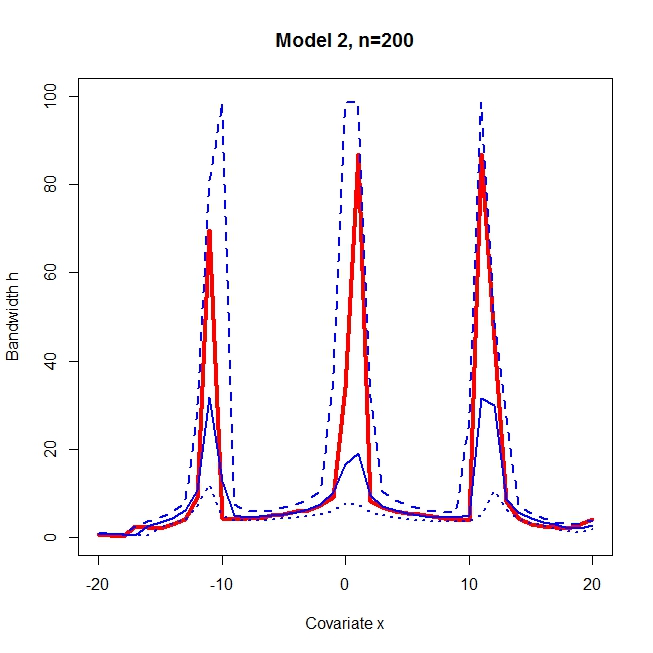} \newline
\caption{Optimal $h_{x,MSE}$ (red line), median (solid blue line), 25th (dotted blue line) and 75th (dashed blue line) percentiles of the bootstrap bandwidth, $h_x^*$, along $m = 1000$ samples of sizes $n = 50$ (top), $n = 100$ (center) and $n = 200$ (bottom), for Model 1 (left) and Model 2 (right).
}
\label{bootstrap_h}
\end{figure}
\thispagestyle{empty}
\clearpage 
}

\section{Application to real data}
\label{sec:5} 

\nolinenumbers

We applied both the semiparametric and the nonparametric
estimators to a real dataset of $414$ colorectal cancer patients from CHUAC
(Complejo Hospitalario Universitario de A Coru\~{n}a), Spain. We considered
two covariates: the stage, from $1$ to $4$, and the age, from $23$ to $103$. The variable $Y$ is the follow-up time (months) since the diagnostic until death. About $50\%$ of the observations are censored, with the percentage of censoring varying from $30\%$ to almost $71\%$, depending on the stage. In Table 1 we show a summary of the data set.

\begin{center}
\begin{tabular}{cccc}
\multicolumn{4}{c}{Table 1: Colorectal cancer patients from CHUAC} \\
Stage & Number of patients & Number of censored data & \% Censoring \\ \hline
1 & 62 & 44 & 70.97 \\
2 & 167 & 92 & 55.09 \\
3 & 133 & 53 & 39.85 \\
4 & 52 & 16 & 30.77 \\ \hline
& 414 & 205 & 49.52%
\end{tabular}
\end{center}

The incidence is estimated with both the semiparametric and the nonparametric estimators. The age of the patients has been considered as a continuous covariate, and the data have been split into four groups according to the categorical covariate stage.

\begin{figure}[tbp]
\includegraphics[width=0.5\textwidth]{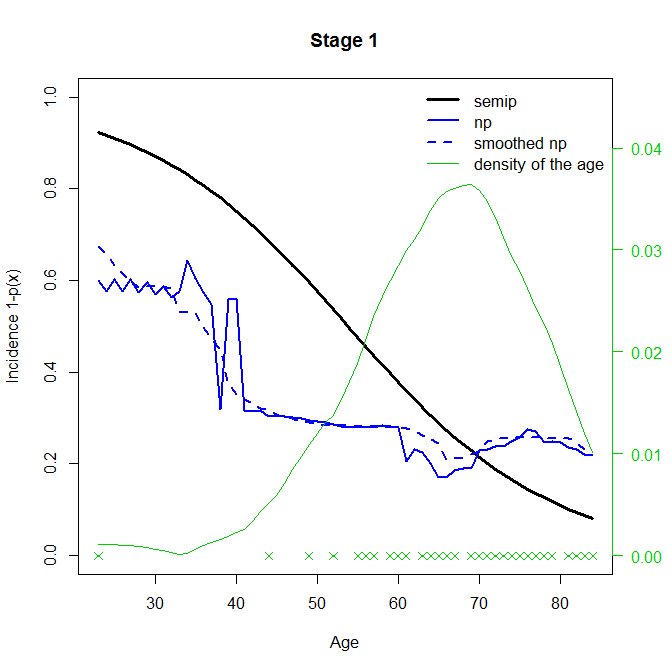} %
\includegraphics[width=0.5\textwidth]{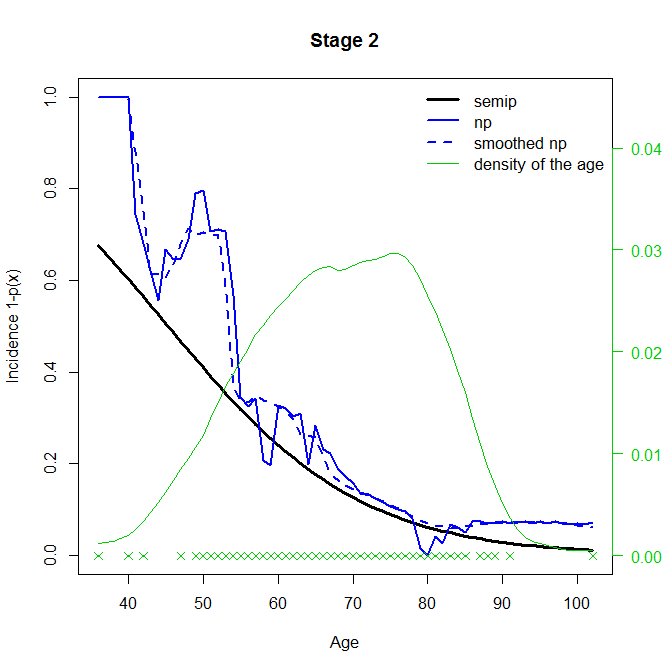} \newline
\includegraphics[width=0.5\textwidth]{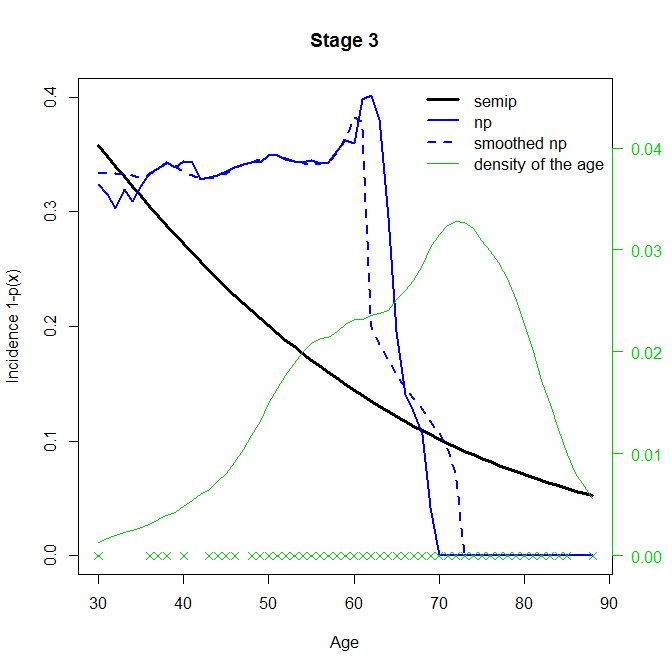} %
\includegraphics[width=0.5\textwidth]{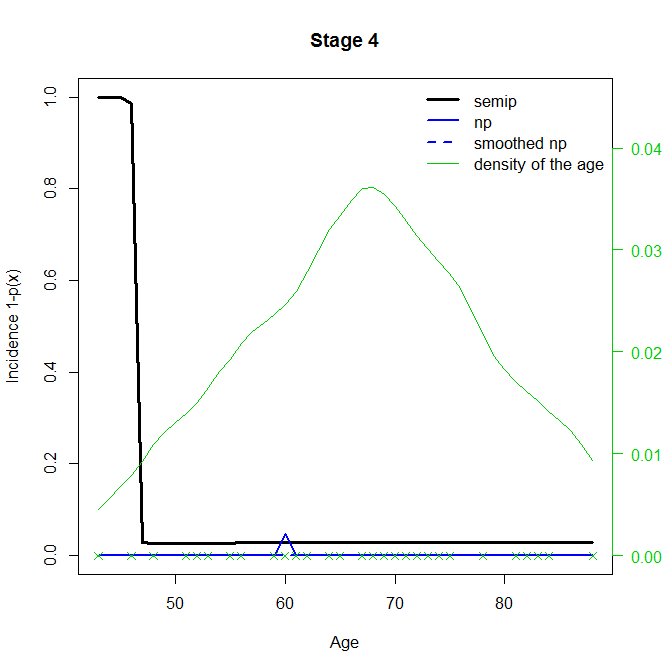}
\caption{Semiparametric (black line) and nonparametric estimations of
the incidence in
Stages 1-4 depending on the age computed with the boots\-trap
bandwidth $h^*_{x}$ (solid blue line) and with the smoothed
bootstrap bandwidth $h^{*\;smoothed}_{x}$ (dashed blue line). The
green line represents the Parzen-Rosenblatt density estimations of the
covariate age, using Sheather and Jones' plug-in bandwidth.} \label{incidence_stage1}
\end{figure}

For the nonparametric estimator of $1-p$, a naive pilot bandwidth selector has been proposed in (\ref{eq:g_dat_sim}) if the distribution of $X$ is uniform. The idea is to provide a data-driven pilot bandwidth which only depends on both the sample size and on the distribution of the covariate, keeping the $n^{-1/9}$ optimal order. Taking into account that in this case the distribution of the covariate is not uniform (see Figure \ref{incidence_stage1}), we propose to use the following local pilot bandwidth:
\begin{equation*} 
g_x = \frac{d_{k}^+(x) + d_{k}^-(x)}{2}100^{1/9} n^{-1/9},
\end{equation*}
where $d_k^+(x)$  is the distance from $x$ to the $k$-th nearest neighbor on
the right, $d_k^-(x)$ the distance from $x$ to the $k$-th nearest neighbor
on the left, and $k$ a \mbox{suitable} integer depending on the sample size. If
there are not at least $k$ neighbors on the right (or left), we use $%
d_k^+(x)=d_k^-(x)$ (or $d_k^-(x)=d_k^+(x)$) res\-pec\-ti\-vely. Our numerical experience shows that a good choice is to consider $k = n/4$. Note that when $n = 100$ the value of the local pilot bandwidth $g_x$ is the mean distance to the $25$th nearest neighbor on both the left and right sides.

For the nonparametric estimator, alongside the bootstrap bandwidth, we
have also used a smoothed bootstrap bandwidth. We followed \cite{Cao3}, who applied a method for smoothing local bandwidths in another context. 
The bootstrap bandwidths have been computed in the equispaced grid $x_0 < x_1 < \dots < x_m$ of the interval $[X_{(1)},X_{(n)}]$ given by $x_i = X_{(1)}+\Delta i,i=0,1,2,\dots, m$ where $\Delta = (X_{(n)}-X_{(1)})/m$. The smoothed bootstrap bandwidth in point $x_l$ is computed as follows:
\begin{equation*}
h_{x_l}^{*\;smooth} =
\begin{cases}
\frac{\sum_{j=0}^{l+5} h_{x_j}^*}{l+6}, & l=0, 1, 2, 3, 4 \\
\frac{\sum_{j=l-5}^{l+5} h_{x_j}^*}{11}, & l=5, 6, 7, \dots, m-5 \\
\frac{\sum_{j=l-5}^{m} h_{x_j}^*}{m-l+6}, & l=m-4, m-3, m-2, m-1, m
\end{cases} .
\end{equation*}
Figure \ref{incidence_stage1} shows the estimations of the
probability of being cured for the different stages with respect to
the age of the patients. We can see that the effect of the age on
the incidence changes with the stage. For example, using the nonparametric incidence estimation, in Stage $1$, patients have a probability of survival between 25\% and $65\%$, depending on the age; whereas in
Stage $3$, for patients above $60$, in a 10 years gap that
probability decreases considerably from 40\% to almost 0\%. It is
important to highlight the difference between the nonparametric and
the semiparametric curves, that seems to indicate that the logistic
model is not valid for the data. The results in Stage $4$ deserve
some comments. A total of 11 in the 12 greatest lifetimes in Stage
$4$, including the largest lifetime, are uncensored and,
consequently, uncured. This causes that the nonparametric estimation
of the probability of being cured is equal to $0$. Although it
should not be stated that it is impossible for a patient with Stage
$4$ colorectal cancer to survive, this estimation reinforces the
assertion that long-term survival in patients with Stage $4$
colorectal cancer is uncommon \citep{Miyamoto}. This fact, far from
being a weakness of the nonparametric method, is an important
advantage, since it allows to detect situations in which introducing
the possibility of cure does not contribute to improve the model.

\begin{figure}[tbp]
\includegraphics[width=0.5\textwidth]{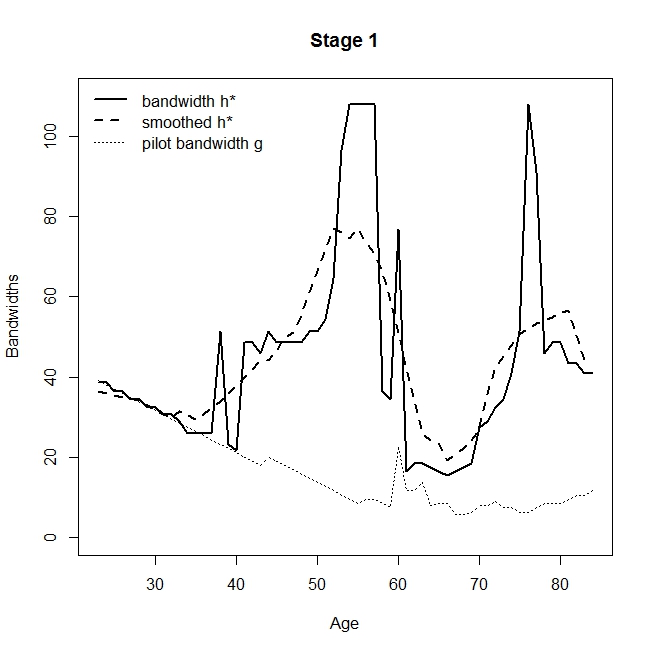}
\includegraphics[width=0.5\textwidth]{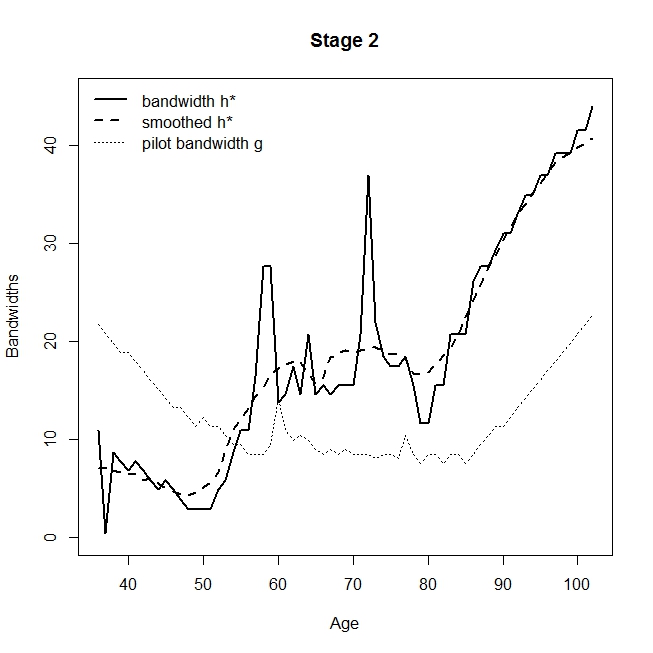} \newline
\includegraphics[width=0.5\textwidth]{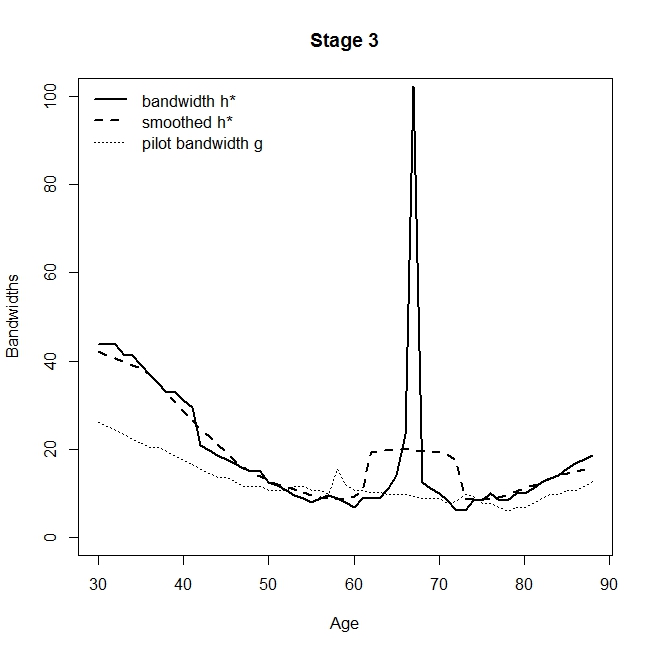}
\includegraphics[width=0.5\textwidth]{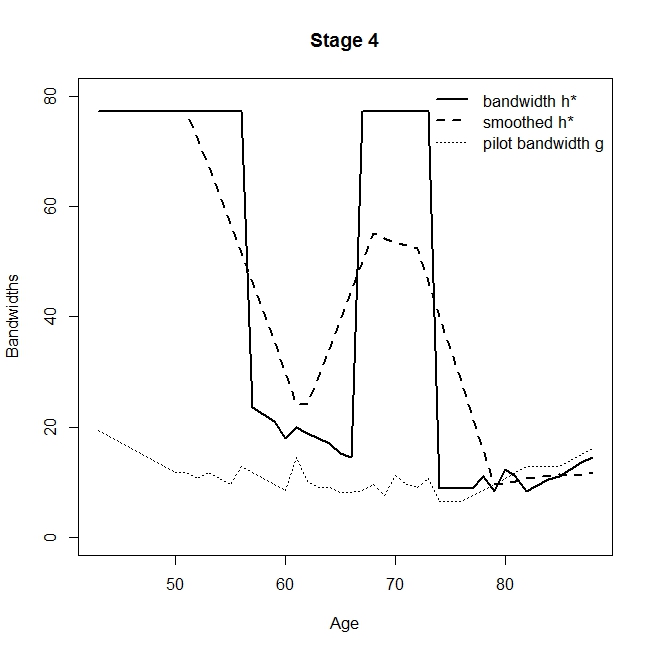}
\caption{Bootstrap bandwidth $h^{*}_{x}$ (solid line), smoothed bootstrap bandwidth $h^{*\;smoothed}_{x}$ (dashed line) and local pilot bandwidth $g_x$ (dotted line) used for the nonparametric incidence estimator for patients in Stages 1-4.}
\label{h_bootstrap}
\end{figure}

Note that in order to obtain the optimal bootstrap bandwidth, $B=1000$ bootstrap resamples are used. In a similar way as we did in Section \ref{subsec:4_2}, we carry out a one-step procedure with a double search. We consider a number of $21$ bandwidths equispaced on a logarithmic scale in both searches. The first search is performed between $0.2$ and the empirical range of $X$. The second one is carried out using another grid centered around the optimal bandwidth obtained in the first search. We show the resulting bootstrap bandwidths, with the corresponding local pilot bandwidths, for the different values of the covariate age in Figure \ref{h_bootstrap}.

\begin{figure}[tbp]
\includegraphics[width=0.5\textwidth]{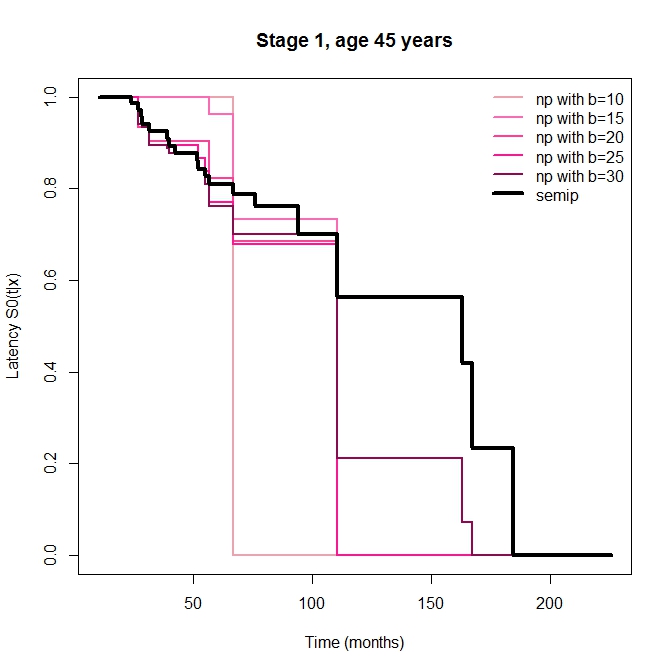} %
\includegraphics[width=0.5\textwidth]{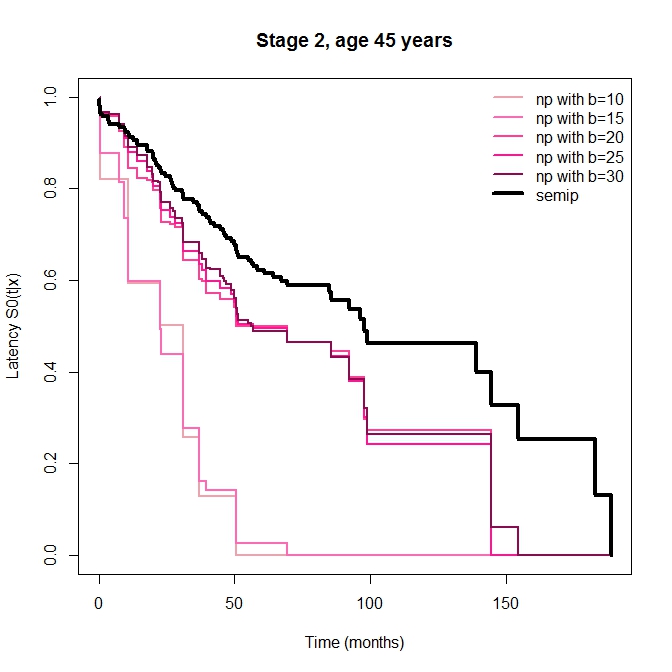} \newline
\includegraphics[width=0.5\textwidth]{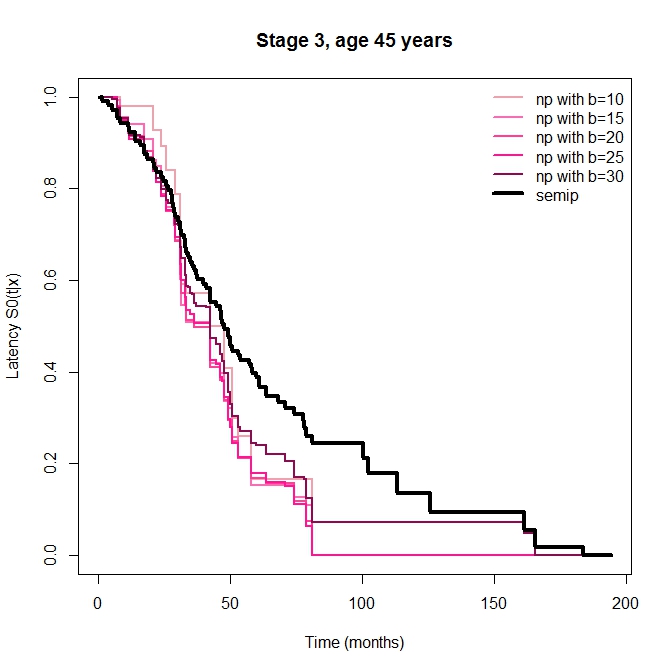} %
\includegraphics[width=0.5\textwidth]{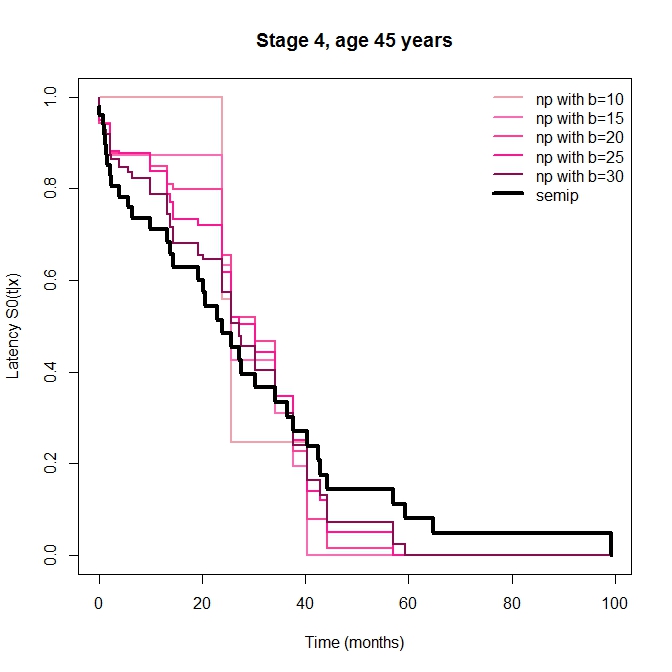}
\caption{Estimated latency for patients of age $45$ in Stages 1-4, using
the semiparametric (black line) and nonparametric estimators with 5 equispaced bandwidths
ranging from $b_{0} = 10$ (light pink line) to $b_{4} = 30$ (dark pink line). }
\label{latency_stage_45}
\end{figure}

\begin{figure}[tbp]
\includegraphics[width=0.5\textwidth]{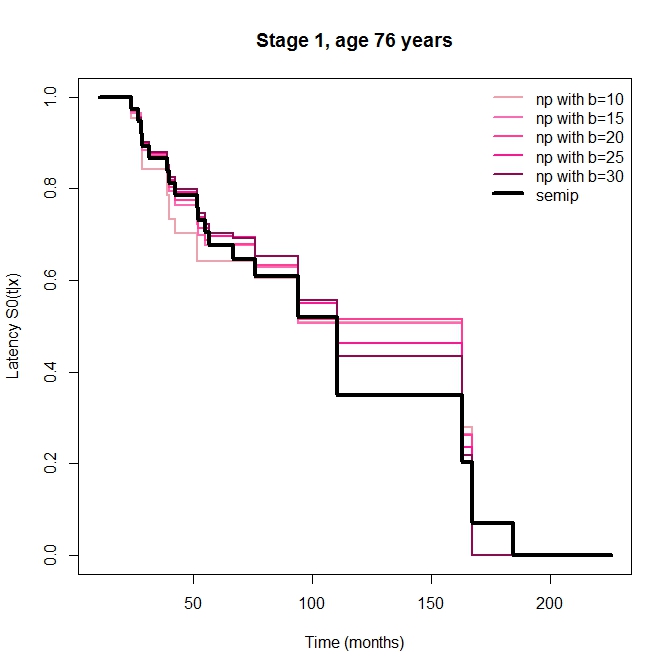} %
\includegraphics[width=0.5\textwidth]{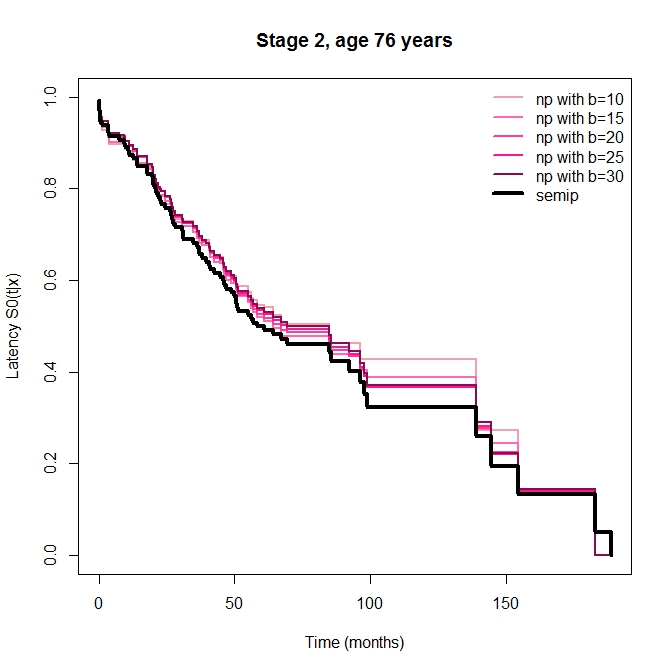} \newline
\includegraphics[width=0.5\textwidth]{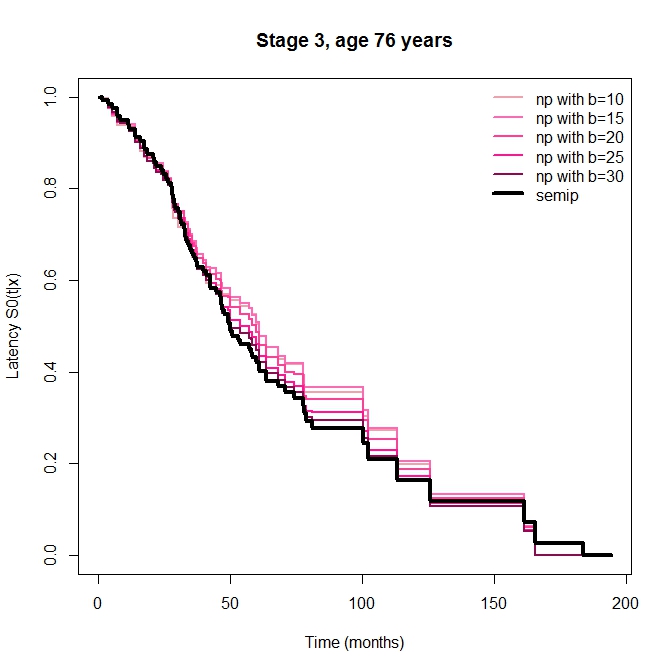} %
\includegraphics[width=0.5\textwidth]{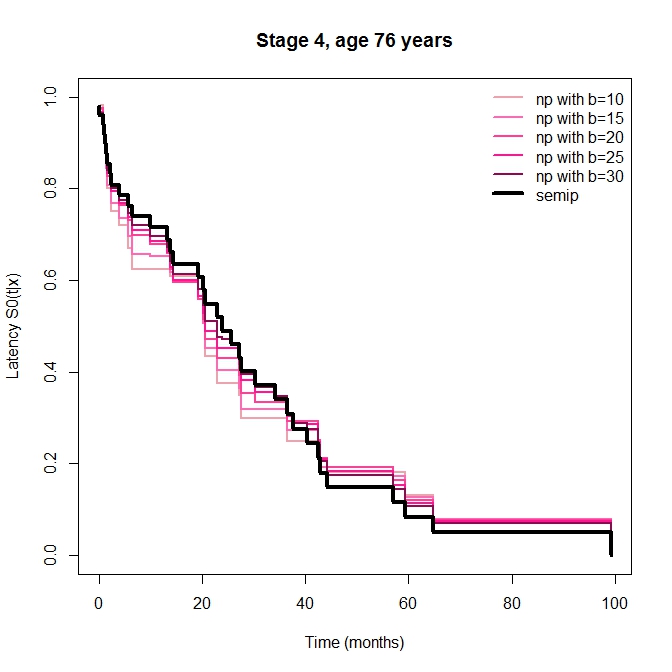}
\caption{Estimated latency for patients of age $76$ in Stages 1-4, using the semiparametric (black line) and nonparametric estimators with 5 equispaced bandwidths ranging from $b_{0} = 10$ (light pink line) to $b_{4} = 30$ (dark pink line).}
\label{latency_stage_76}
\end{figure}

In Figures \ref{latency_stage_45} and \ref{latency_stage_76} we show the
latency estimation for Stages $1,2,3$ and $4$ for two diffe\-rent ages, $45$
and $76$. The nonparametric estimator $\hat S_{0,b_x}$ is computed with five
di\-ffe\-rent constant bandwidths: $b=10,15,20,25$ and $30$. It is noteworthy that in Stages $1$ and $2$ for $45$ years, the bandwidth selection influences considerably latency estimation.
 This is due to the low density of the covariate around this age, as we can see in Figure \ref{incidence_stage1}.
Suggested by the results in Section \ref{subsec:4_1}, it is reasonable to choose a large bandwidth. Nevertheless, the choice of the bandwidth for the latency estimation is out of the scope of this paper. This issue has been very recently studied by \cite{Lopezetal}.

\section*{Appendix}
\nolinenumbers

\begin{proof}[\emph{\textbf{Proof of Theorem \ref{thm:mle}}}] The idea is to estimate $p\left( x\right) $ locally, maxi\-mi\-zing the observed local likelihood function around $x$. It can be proved that the maximum likelihood estimator of the survival function $S_{0}(t|x)=1 - F_0(t|x) $ has jumps only at the observations $\left( X_{i},T_{i},\delta_{i}\right) ,i=1,\ldots ,n$ with jumps%
\begin{equation*}
q_{i}\left( x\right) = S_{0}(T_{(i)}^{-}|x) - S_{0}(T_{(i)}|x).
\end{equation*}
The local likelihood of the model is%
\begin{eqnarray*}
 L\left( p\left( x\right) ,S_{0}\left( \cdot |x\right) \right) =&  \prod\limits_{i=1}^{n}\left\{ \left[ p\left( x\right) q_{i}\left( x\right) \right]
^{B_{h\left( i\right) }\left( x\right) \delta _{(i)}}\left[ 1-p\left( x\right)
+p\left( x\right) \right. \right. \\
&  \times   \left. \left. \left( 1-\sum_{j=1}^{i-1}q_{j}\left( x\right) \right) %
\right] ^{\left( 1-\delta _{(i)}\right) B_{h\left( i\right) }\left( x\right)
}\right\} .
\end{eqnarray*}%
Let $D_{i}\left( x\right) =B_{h\left( i\right) }\left( x\right) \delta _{(i)}$
and $P_{i}\left( x\right) =p\left( x\right) q_{i}\left( x\right) $, then%
\begin{equation*}
L\left( p\left( x\right) ,S_{0}\left( \cdot |x\right) \right)
=\prod_{i=1}^{n}\left\{ P_{i}\left( x\right) ^{D_{i}\left( x\right) }\left(
1-\sum_{j=1}^{i-1}P_{j}\left( x\right) \right) ^{B_{h\left( i\right) }\left(
x\right) -D_{i}\left( x\right) }\right\} .
\end{equation*}%
Consider now the functions $\lambda _{i}\left( x\right) =P_{i}\left(
x\right) /\left( 1-\sum\limits_{j=1}^{i-1}P_{j}\left( x\right) \right) $
satisfying%
\begin{equation}
1-\sum_{j=1}^{k}P_{j}\left( x\right) =\prod_{j=1}^{k}(1-\lambda _{j}\left(
x\right) ).   \tag{A.1} \label{res2}
\end{equation}%
Straightforward calculations yield
\begin{equation*}
L\left( \lambda _{1}(x),\dots ,\lambda _{n}(x) \right)
=\prod_{i=1}^{n}\lambda _{i}\left( x\right) ^{D_{i}\left( x\right) }\left(
1-\lambda _{i}\left( x\right) \right) ^{\sum\limits_{r=i+1}^{n}B_{h\left(
r\right) }\left( x\right) }.
\end{equation*}%

Maximizing the likelihood of the observations for the cure model is
equivalent to \mbox{maximizing}%
\begin{equation*}
\max_{\lambda _{i}\geq 0;i=1,\dots ,n}\Psi (\lambda _{1},\dots ,\lambda _{n}),
\end{equation*}%
where $\Psi $ is the local loglikelihood:
\begin{equation*}
\Psi (\lambda _{1}\left( x\right) ,\dots ,\lambda
_{n}\left( x\right) )=\sum_{i=1}^{n}\left[ D_{i}\left( x\right) \log \lambda
_{i}\left( x\right) +  \left ( \sum\limits_{r=i+1}^{n}B_{h\left( r\right) }\left(
x\right)  \right ) \log \left( 1-\lambda _{i}\left( x\right) \right) \right],
\end{equation*}%
subject to
\begin{equation}
\prod_{i=1}^{n}(1-\lambda _{i}\left( x\right) )=1-\sum_{j=1}^{n}P_{j}\left(
x\right) =1-p\left( x\right) .  \tag{A.2} \label{res}
\end{equation}%
Using standard maximization techniques, we obtain
\begin{equation*}
\widehat{\lambda }_{i}\left( x\right) =\frac{D_{i}\left( x\right) }{%
\sum\limits_{r=i+1}^{n}B_{h\left( r\right) }\left( x\right) +D_{i}\left(
x\right) }=\frac{\delta _{(i)}B_{h\left( i\right) }\left( x\right) }{%
\sum\limits_{r=i+1}^{n}B_{h\left( r\right) }\left( x\right) +\delta
_{(i)}B_{h\left( i\right) }\left( x\right) }.
\end{equation*}%
Replacing $\lambda_i$ in (\ref{res}) by $\widehat {\lambda}_i(x)$, we obtain the estimator of $1 - p\left( x\right) $ given in (\ref{ec:p_estimation}).

With respect to the distribution of the uncured subjects, note that%
\begin{equation*}
F_{0}\left( T_{(i)}|x\right) =\sum_{j=1}^{i}q_{j}\left( x\right).
\end{equation*}%
Since the jumps satisfy $P_{i}\left( x\right) =p\left( x\right) q_{i}\left( x\right) $
and using (\ref{res2}), we find that the local maximum likelihood estimator is given by
\begin{equation*}
\hat{F}_{0}\left( T_{(i)}|x\right) =\frac{1}{\hat{p}_h \left( x\right) }\left[ 1-\prod_{j=1}^{i}\left( 1-\hat{\lambda }_{j}\left( x\right) \right) \right] =\frac{\hat{F_h}\left( T_{(i)}|x\right) }{\hat{p}_h\left( x\right) },
\end{equation*}
with $\hat{F_h}\left( T_{(i)}|x\right) $ the Beran estimator of $F=1-S$ computed at time $T_{(i)}$.
\end{proof}
\bigskip

The following auxiliary results are necessary to prove Theorem \ref{thm:iid}.

\begin{lem}[Xu and Peng (2014)]
\label{Th_Tn_conv_tau0} Under assumption (A10),
\begin{equation*}
T_{\max}^{1}=\max_{i:\delta _{i}=1}\left( T_{i}\right) \rightarrow \tau _{0}%
\text{ in probability as }n\rightarrow \infty \text{.}
\end{equation*}
\end{lem}

\begin{lem}
\label{Th_png} Under assumption (A9), we have that
\begin{equation*}
n^{\alpha} (\tau_0 -T_{\max}^{1}) \rightarrow 0 \text{ \ a.s.}
\end{equation*}
for any $\alpha \in (0,1)$. In particular, for a sequence of bandwidths
satisfying $nh^5(\ln n)^{-1}=O(1)$, we have
\begin{equation*}  \label{eq:png10}
\tau_0 - T_{\max}^{1} = o\left ( \left ( \frac{\ln n}{nh} \right )^{3/4}
\right ) \text{a.s.}  \tag{A.3}
\end{equation*}
\end{lem}

\begin{proof}[\emph{\textbf{Proof of Lemma \ref{Th_png}}}] Using the Borel-Cantelli lemma, it is sufficient to prove that
\begin{equation}
\label{eq:png44}
\sum_{n=1}^{\infty} P \left ( |a_n (\tau_0 -  T_{\max}^{1}) | > \epsilon    \right )  < \infty, \text{ for all } \epsilon > 0, \tag{A.4}
\end{equation}
where $a_n = n^{\alpha}$. Let us fix $\epsilon > 0 $ and consider:
\begin{flalign}
& P(|a_n (\tau_0 -  T_{\max}^{1}) | > \epsilon) \notag \\
 =& P \left ( T_{\max}^{1} < \tau_0 - \frac{\epsilon}{a_n} \right ) \notag\\
=& P \left (T_{i} < \tau_0 - \frac{\epsilon}{a_n}, \text{ for all } i=1, 2, \dots n \text{ where } \delta_i=1 \right ) \notag  \\
=& E \left [ P \left ( T_i < \tau_0 - \frac{\epsilon}{a_n},  \text{ for all } i=1, 2, \dots n \text{ where } \delta_i=1 \big |\delta_1, \delta_2, \dots, \delta_n \right )   \right ]  \notag\\
=& E \left [ \prod_{i=1}^n  P \left ( T_i < \tau_0 - \frac{\epsilon}{a_n} \big  |\delta_i=1  \right )^{\delta_i}   \right ] =E \left [ P \left ( T_1 < \tau_0 - \frac{\epsilon}{a_n} \big  |\delta_1=1  \right )^{\sum_{i=1}^n \delta_i}   \right ] \notag\\
=&  E \left [ \left ( H_{c,1}  \left ( \tau_0 - \frac{\epsilon}{a_n} \right ) \right )^{\sum_{i=1}^n \delta_i}  \right ] \notag,
\end{flalign}
where
\begin{displaymath}
H_{c,1} (t) = P \left ( T < t \big |\delta=1 \right ) = \frac{P(T < t, \delta=1)}{P(\delta=1)} = \frac{H_1(t)}{\rho},
\end{displaymath}
with $\rho = P(\delta=1) = E(\delta)$ and $H_1(t)=P(T<t, \delta=1)$. Consequently, since \mbox{$\sum_{i=1}^n \delta_i \stackrel{\text{d}}{=} B(n,\rho) $}, we get:
\begin{flalign}
\label{eq:png1}
& P(|a_n (\tau_0 -  T_{\max }^{1}) | > \epsilon)   \tag{A.5} \\
=& E \left [ H_{c,1}  \left ( \tau_0 - \frac{\epsilon}{a_n} \right )^{\sum_{i=1}^n \delta_i}  \right ] = \sum_{j=0}^n \binom{n}{j} \rho^j (1 - \rho)^{n-j} H_{c,1}  \left ( \tau_0 - \frac{\epsilon}{a_n} \right )^j \notag \\
=&\sum_{j=0}^n \binom{n}{j} \left [ \rho  H_{c,1}  \left ( \tau_0 - \frac{\epsilon}{a_n} \right ) \right ]^j (1 - \rho)^{nj} = \left [ \rho  H_{c,1}  \left ( \tau_0 - \frac{\epsilon}{a_n} \right ) + 1 - \rho \right ]^n  \notag \\
=& \left [ \rho \left ( H_{c,1}(\tau_0)  - \frac{\epsilon}{a_n} H^{'}_{c,1}(\tau_0) + \frac{\epsilon^2}{2 a_n^2} H^{''}_{c,1}(\xi_n) \right ) + 1 - \rho \right ]^n \notag \\
=& \left [ \rho - \rho  \frac{\epsilon}{a_n} H^{'}_{c,1}(\tau_0) + \rho \frac{\epsilon^2}{2 a_n^2} H^{''}_{c,1}(\xi_n) + 1 - \rho \right ]^n \notag \\
=& \left ( 1 - \rho  \frac{\epsilon}{a_n} H^{'}_{c,1}(\tau_0) + \rho \frac{\epsilon^2}{2 a_n^2} H^{''}_{c,1}(\xi_n)  \right )^n, \notag
\end{flalign}
for some $\xi_n \in \left [ \tau_0 - \frac{\epsilon}{a_n}, \tau_0 \right ]$, since $H_{c,1}(\tau_0)=1$. \\
	
Using assumption (A9), $\sup_{t \geq 0} |H^{''}_{c,1}(t)| = C < \infty $. As a consequence, since $\epsilon/a_n \rightarrow 0$ as $n\rightarrow \infty$, then there exists some $n_0 \in \Bbb N $ such that for all $n \geq n_0$:
\begin{equation}
\label{eq:png2}
\bigg | \rho  \frac{\epsilon^2}{2 a_n^2} H^{''}_{c,1}(\xi_n) \bigg | \leq  \frac{\rho \epsilon^2}{2 a_n^2} C \leq \rho \frac{\epsilon}{2 a_n}  H^{'}_{c,1}(\tau_0). \tag{A.6}
\end{equation}

From (\ref{eq:png1}) and (\ref{eq:png2}), we have that:
\begin{equation}
P ( |a_n (\tau_0 -  T_{\max}^{1} ) | > \epsilon ) \leq  \left ( 1 - \rho \frac{\epsilon}{2 a_n} H^{'}_{c,1}(\tau_0)  \right )^n = \left ( 1 - \frac{\epsilon}{2 a_n} H^{'}_{1}(\tau_0)  \right )^n = b_n^{n/a_n}, \tag{A.7} \label{eq:png3}
\end{equation}
where
\begin{equation}
\label{eq:png4}
b_n = \left ( 1 - \frac{\epsilon}{2 a_n} H^{'}_{1}(\tau_0)   \right )^{a_n} \xrightarrow[n \rightarrow \infty]{} r, \tag{A.8}
\end{equation}
with $r = \exp \left (- \frac{ \epsilon H^{'}_{1} (\tau_0)}{2} \right ) < 1$.

Using (\ref{eq:png3}) and (\ref{eq:png4}), to prove (\ref{eq:png44}) it suffices to show that $\sum_{n=1}^{\infty} r^{n/a_n} < \infty $. For that purpose, we will prove that
\begin{equation}
\label{eq:png5}
r^{n/a_n} < n^{-2}, \text{ for $n$ large enough} \tag{A.9}
\end{equation}
and, since the hyperharmonic series $\sum_{n=1}^{\infty} n^{-2}$ is convergent, the series $\sum_{n=1}^{\infty} r^{n/a_n}$ will also be convergent.

Note that inequality (\ref{eq:png5}) can be written as
\begin{equation}
\label{eq:png6}
2 \log_R n < \frac{n}{a_n}, \tag{A.10}
\end{equation}
with $R = r^{-1} \in (1, \infty)$. Recall that $a_n = n^{\alpha}$ for some $\alpha \in (0, 1)$. Now condition (\ref{eq:png6}) becomes
\begin{equation*}
2 \log_R n < n^{1- \alpha},
\end{equation*}
which is true for $n$ large enough, since $n^{-(1-\alpha)}2\log_R n \rightarrow 0$. As a consequence, $n^{\alpha} (\tau_0 -  T_{\max }^{1}) \rightarrow 0 \text{ a.s. }$ for any $\alpha \in (0, 1)$. On the other hand, note that:
\begin{eqnarray*}
 \frac{n^{-\alpha}}{ \left ( \frac{\ln n}{n h} \right )^{3/4} } = \left [ \frac{nh^5}{\ln n}  \frac{n^{4 - 20\alpha / 3}}{(\ln n)^4} \right ]^{3/20} \xrightarrow[n \rightarrow \infty]{} 0,
\end{eqnarray*}
for $\alpha \geq 3/5$ and a sequence of bandwidths satisfying $(\ln n)^{-1}nh^5=O(1)$. Therefore, the result in (\ref{eq:png10}) holds. This completes the proof.
\end{proof}

In the next three lemmas, we use existing results in the literature for a
fixed $t$ such that $1 - H(t|x) \geq \theta > 0$ in $(t,x) \in [a,b] \times
I_{\delta}$, and apply them to the random value $t= T_{\max }^{1}$. Note that if $%
\tau_0 < \tau_G(x) = \tau_H(x)$ for all $x \in I_{\delta}$, then from Lemma \ref{Th_Tn_conv_tau0}, under assumption (A10), we have that:
\begin{equation*}
T_{\max }^{1} = \max_{i:\delta_i=1}(T_i) \rightarrow \tau_0 < \tau_H(x) \text{
in probability as } n \rightarrow \infty.
\end{equation*}
Therefore, for $n$ large enough, $T_{\max }^{1} \leq \tau_0 < \tau_H(x)$ for
all $x \in I_{\delta}$ and taking $b=\tau_0$ we can apply the results
considering $t= T_{\max }^{1}$.

\begin{lem}
\label{Th_1menosp_exp} Under assumptions (A1)-(A5), (A10) and (A12), and if $nh^{5}/\ln n=O(1)$ and $\ln n/(nh)\rightarrow 0$, then the incidence estimator satisfies:
\begin{equation*}
1-\hat{p}_{h}(x)=\exp \left( -\widehat{\Lambda }_{h}( T_{\max }^{1}|x)\right)
+R_{n} \left( x\right), \text{ for all } x\in I,
\end{equation*}%
with
\begin{equation*}
\sup_{x\in I}\left\vert R_{n}\left( x\right) \right\vert =O\left( \left(
nh\right) ^{-1}\right) \text{a.s.}
\end{equation*}
\end{lem}

\begin{proof}[\emph{\textbf{Proof of Lemma \ref{Th_1menosp_exp}}}] The incidence estimator is equal to:
\begin{equation*}
1-\hat{p}_{h}(x)=1-\hat{F}_{h}( T_{\max }^{1}|x),
\end{equation*}%
where $\hat F_h(t|x)=1 - \hat S_h(t|x)$ is the Beran estimator in (\ref{cheda:S_est}). The result derives directly for $\hat{F}_{h}( t|x)$ from the so-called property 3) in the proof of part c) of Theorem 2 in \cite{Iglesias-Perez}, when the data are subject to random left truncation and right censorship, for which assumptions (A1),(A3)-(A5) and (A12) are required. Assumptions (A2) and (A10) allow to use the aforementioned property when $t= T_{\max }^{1}$.  \cite{Gonzalez-Manteiga} proved a similar result under right random censoring with fixed design on the covariate.
\end{proof}

\begin{lem}
\label{TH_iidL} 
Under assumptions (A1)-(A11) and (A13) for $x\in I$ and if $nh^{5}/\ln n=O(1)$, $\ln n/(nh)\rightarrow 0$, then
\begin{equation*}
\widehat{\Lambda }_{h}( T_{\max }^{1}|x)-\Lambda( T_{\max
}^{1}|x)=\sum_{i=1}^{n}\tilde B_{h i}(x) \xi \left(
T_{i},\delta_{i},x\right) + \tilde{R}_{n} \left( x\right),
\end{equation*}
with $\tilde B_{hi}$ in (\ref{Btilde}), $\xi$ in (\ref{xi}) and
\begin{equation*}
\sup_{x\in I}\left\vert \tilde{R}_{n} \left( x\right) \right\vert =O\left(
\left( \frac{\ln n}{nh}\right) ^{3/4}\right) \text{a.s.}
\end{equation*}
\end{lem}

\begin{proof}[\emph{\textbf{Proof of Lemma \ref{TH_iidL}}}] Under assumptions (A1)-(A8), (A10) and (A13), we apply Theo\-rem 2(b) of \cite{Iglesias-Perez} (similarly Theo\-rem 2.2 of \cite{Gonzalez-Manteiga} with fixed design using GM weights) to $t=T^1_{\max}$:
\begin{flalign*}
\label{A.12}
\widehat{\Lambda}_{h}( T_{\max }^{1}|x)-\Lambda( T_{\max }^{1}|x)
&=\sum_{i=1}^{n} \tilde B_{h i}(x) \xi \left( T_{i},\delta_i ,x \right)
 + \sum_{i=1}^{n}(B_{h i}(x) - \tilde B_{h i}(x)) \xi \left( T_{i},\delta_i ,x \right) \notag \\
&
+\sum_{i=1}^{n}B_{h i }(x)\left(\tilde \xi \left( T_{i},\delta_{i},x\right)-\xi \left( T_{ i},\delta_{ i},x\right)\right) + \tilde{\tilde{R}}_{n}\left( x\right),  \tag{A.11}
\end{flalign*}
with $\xi$ in (\ref{xi}),
\begin{equation*}
\tilde \xi \left(T_{i},\delta _{i},x\right) =\frac{I(\delta _{i}=1)}{1-H(T_i|x)}-\int_{0}^{ T_{\max}^{1}}\frac{I(t < T_{i})}{\left( 1-H(t|x \right)) ^{2}}dH^{1}(t|x)
\end{equation*}
and
\begin{equation*}
\sup_{x\in I}\left\vert \tilde{\tilde{R}}_{n} \left( x\right) \right\vert =O\left( \left(
\frac{\ln n}{nh}\right) ^{3/4}\right) a.s.
\end{equation*}
Note that
\begin{equation*}
\left \vert \tilde \xi \left( T_{i},\delta_{i},x\right)-\xi \left( T_{i},\delta_{i},x\right) \right \vert
\leq \int_{ T_{\max }^{1}}^{\tau_0}\dfrac{dH^{1}(t|x)}{\left( 1-H(t^{-}|x \right)) ^{2}} \text{ \ \ } \text{for all } i=1,\ldots,n.
\end{equation*}

Then, under assumption (A9) we apply Lemma \ref{Th_png}, and assuming (A11), it is easy to prove that for a sequence of bandwidths satisfying $nh^5(\ln n)^{-1}=O(1)$, the third term in (\ref{A.12}) is,
\begin{equation*}
\sup_{x\in I}\left\vert\sum_{i=1}^{n}B_{h i }(x)\left(\tilde \xi \left( T_{i},\delta_{i},x\right)-\xi \left( T_{ i},\delta_{ i},x\right)\right)\right\vert = o\left(\left(\frac{\ln n}{nh} \right)^{3/4} \right)a.s.
\end{equation*}

For the second term in (\ref{A.12}), it is important to note that:
\begin{eqnarray*}
\sum_{i=1}^n(B_{h i}(x) - \tilde B_{h i}(x)) \xi(T_i,\delta_i,x)  &=& \frac{1}{nh} \sum_{i=1}^n K \left ( \frac{x - X_i}{h} \right ) \xi(T_i,\delta_i,x) {\frac{m(x) - \hat m_{h}(x)}{\hat m_{h}(x) m(x)}},
\end{eqnarray*}
with $\hat m_h(x)$ the Parzen-Rosenblatt estimator of $m(x)$. Using Theorem 3.3 of \cite{Arcones}, standard bias and variance calculations and Taylor expansions lead to
\begin{displaymath}
\sup_{x \in I} \left \vert \frac{1}{nh}\sum_{i=1}^n K \left ( \frac{x - X_i}{h} \right )\xi(T_i,\delta_i,x) \right \vert =O \left ( h^2 + \sqrt{\frac{\ln \ln n}{nh}} \right ) \text{a.s.}
\end{displaymath}
Using again Theorem 3.3 of \cite{Arcones}, it is easy to prove that:
\begin{equation*}
\sup_{x\in I} \left \vert \frac{m(x)-\hat m_{h}(x)}{\hat m_{h}(x) m(x)} \right \vert = O \left ( h^2 + \sqrt{\frac{\ln \ln n}{nh}} \right ) \text{ a.s.}
\end{equation*}
Therefore,
\begin{equation*}
\sup_{x\in I} \left \vert \sum_{i=1}^n(B_{h i}(x) - \tilde B_{h i}(x))\xi(T_i,\delta_i,x)
 \right \vert = O \left (\left ( h^2 + \sqrt{\frac{\ln \ln n}{nh}} \right )^2 \right ) \text{ a.s.}
\end{equation*}
For a sequence of bandwidths satisfying $nh^5(\ln n)^{-1}=O(1)$, it is immediate to prove that
\begin{equation*}
\sup_{x\in I} \left \vert \sum_{i=1}^n(B_{h i}(x) - \tilde B_{h i}(x))\xi(T_i,\delta_i,x)
\right \vert = O \left ( \left ( \frac{\ln n}{n h} \right )^{3/4} \right ) \text{ a.s.}
\end{equation*}
This completes the proof.
\end{proof}

\begin{lem}
\label{Th_Hn_menos_H} Under assumptions (A1)-(A8), (A10), (A12) and (A13) and if $nh^{5}/\ln n=O(1)$, $\ln n/(nh)\rightarrow 0$, then
\begin{equation*}
\sup_{x\in I}\left\vert \widehat{\Lambda}_{h}\left( T_{\max }^{1}|x\right) -
\Lambda \left(  T_{\max }^{1}|x\right) \right\vert = O\left( \left( \frac{\ln
n}{nh}\right)^{1/2}\right) \text{ a.s.}
\end{equation*}
\end{lem}

\begin{proof}[\emph{\textbf{Proof of Lemma \ref{Th_Hn_menos_H}}}] The equivalent result for a fixed $t \in [a,b]$ is within pro\-per\-ty 2) in the proof of part c) of Theorem 2 in \cite{Iglesias-Perez}, for which assumptions (A1), (A3)-(A8), (A12) and (A13) are required. Assumptions (A2) and (A10) are needed to apply that result to $t=T_{\max}^1$. For the uniform strong consistency of the Beran estimator $\hat F_h(t|x)$, see also \cite{Dabrowska1}.
\end{proof}

\begin{proof}[\emph{\textbf{Proof of Theorem \ref{thm:iid}}}] The incidence estimator can be split into the following terms:
\begin{flalign*}
\label{iid1}
& \left( 1-\hat{p}_h(x)\right) -\left( 1-p(x) \right) \notag \\
& = \hat{S_h}( T_{\max }^{1}|x)-\left( 1-p(x) \right) \notag \\
& = \exp \left[ -\widehat{\Lambda}_h( T_{\max }^{1}|x)\right] -\exp \left[ -\Lambda ( T_{\max }^{1}|x)\right] + R_2(x) + R_3(x), \tag{A.12}
\end{flalign*}
with
\begin{flalign*}
R_2(x) & = \hat{S}_h( T_{\max }^{1}|x) - \exp \left[ -\widehat{\Lambda}_h( T_{\max }^{1}|x)\right]
\end{flalign*}
and
\begin{flalign*}
R_3(x) & = S( T_{\max }^{1}|x)-\left( 1-p(x) \right).
 \end{flalign*}
To the first term of (\ref{iid1}) we apply a Taylor expansion of the function $\exp(y)$ around the value $y=-\Lambda ( T_{\max }^{1}|x)$:
\begin{flalign*}
\label{iid3}
&\exp \left[ -\widehat{\Lambda}_h( T_{\max }^{1}|x)\right]  - \exp
\left[ -\Lambda ( T_{\max }^{1}|x)\right]   \\
&= - \exp\left[ -\Lambda ( T_{\max }^{1}|x)\right] \left( \widehat{\Lambda}_h( T_{\max }^{1}|x)-\Lambda ( T_{\max }^{1}|x)\right) + R_1(x),
\end{flalign*}
with
\begin{equation*}
R_1(x) =  \frac{1}{2}\exp\left[-\Lambda ^{\ast }( T_{\max }^{1}|x)\right] \left( \widehat{\Lambda
}_h( T_{\max }^{1}|x)-\Lambda ( T_{\max }^{1}|x)\right)^{2}
\end{equation*}
and $\Lambda ^{\ast }( T_{\max }^{1}|x)=\eta _{n}\left( x\right)$ a value between $\widehat{\Lambda}_h( T_{\max }^{1}|x)$ and $\Lambda ( T_{\max }^{1}|x)$. Now, adding and substracting $1-p(x)$, and bearing in mind that $S( T_{\max }^{1} | x) = \exp [ - \Lambda( T_{\max }^{1}) | x ] $,
\begin{flalign*}
\label{iid3}
&\exp \left[ -\widehat{\Lambda}_h( T_{\max }^{1}x)\right]  - \exp
\left[ -\Lambda ( T_{\max }^{1}|x)\right]   \\
&= \left( 1-p\left( x\right) \right) \left( \widehat{\Lambda}_h( T_{\max }^{1}|x)-\Lambda ( T_{\max }^{1}|x)\right) + R_1(x) + R_4(x),  \tag{A.13} 
\end{flalign*}
where
\begin{equation*}
R_4(x) = \left[ S\left( T_{\max }^{1}|x\right) -\left( 1-p\left( x\right) \right) \right] \left(\widehat{\Lambda}_h( T_{\max }^{1}|x)-\Lambda ( T_{\max }^{1}|x)\right).
\end{equation*}

Now, inserting (\ref{iid3}) in (\ref{iid1}), we have:
\begin{flalign}
&\left( 1-\hat{p}_h(x)\right) -\left( 1-p\left( x\right) \right)  \notag \\
&=\left( 1-p\left( x\right) \right) \left( \widehat{\Lambda}_h( T_{\max }^{1}|x)-\Lambda ( T_{\max }^{1}|x)\right) +R_{1}(x)+R_{2}(x)+R_{3}(x)+R_{4}(x). \tag{A.14}
\label{iid_step1}
\end{flalign}

The iid representation of $1-\hat{p}_h(x)$ now follows, assuming (A1)-(A11) and (A13), from Lemma \ref{TH_iidL}.

Let us study the remainder terms in (\ref{iid_step1}), starting with $R_{1}(x) $. Taking into account that $\exp \left[ -\Lambda ^{\ast }( T_{\max }^{1}|x)\right] $ is bounded for all $x\in I$, and applying Lemma \ref{Th_Hn_menos_H}, under the assumptions (A1)-(A8), (A10), (A12) and (A13), we have
\begin{equation*}
\sup_{x\in I}\left\vert R_{1}(x) \right\vert =O\left( \frac{\ln n}{nh}\right) \text{ \ a.s.}
\end{equation*}

Regarding $R_{2}(x) $, under the assumptions (A1),(A3)-(A5), (A10) and (A12), directly from Lemma \ref{Th_1menosp_exp} and using $\ln n / (nh) \rightarrow 0$ we obtain:
\begin{equation*}
\sup_{x\in I}\left\vert R_{2}(x) \right\vert =O\left( \left(
nh\right) ^{-1}\right) =o\left( \left( \frac{\ln n}{nh}\right) ^{3/4}\right)
\text{ \ a.s.}
\end{equation*}%

Focusing on $R_{3}(x)$, note that it can be bounded as follows:
\begin{flalign}
\label{eq:png12}
\sup_{x \in I}\big | R_{3} (x) \big |
& = \sup_{x \in I}\big | S( T_{\max }^{1}|x)-\left( 1-p(x) \right) \big |  \notag \\
&= \sup_{x \in I}\bigg | \left[ \left( 1-p(x) \right) +p(x)S_{0}( T_{\max }^{1}|x)\right]
   -\left( 1-p(x) \right) \bigg | \notag \\
&= \sup_{x \in I} \big | p(x) S_{0}( T_{\max }^{1}|x) \big |
   \leq \sup_{x \in I}\big |  S_{0}( T_{\max }^{1}|x)\big | \notag \\
&  = \sup_{x \in I} \big |  S_{0}( T_{\max }^{1}|x) - S_{0}(\tau_0|x)\big | \notag\\
&\leq \sup_{x \in I} | ( T_{\max }^{1}- \tau_0) S_0'(\tau_n | x)|,  \tag{A.15}
\end{flalign}
with $\tau_n \in [ T_{\max }^{1}, \tau_0]$. From condition (A6), that implies that there exists some $\lambda > 0$ such that $\sup_{(t,x) \in [a,b]\times I} |S_0'(t|x)| \leq \lambda $, and using (\ref{eq:png10}) and (\ref{eq:png12}) for a sequence of bandwidths satisfying $nh^5(\ln n)^{-1}=O(1)$, we have that:
\begin{displaymath}
\sup_{x \in I} |R_3(x)| = o \left ( \left (  \frac{\ln n}{nh} \right )^{3/4} \right ) \text{a.s.}
\end{displaymath}
Finally, from Lemma \ref{Th_Hn_menos_H}, the term $R_{4}$ is negligible with respect to $R_3$, and therefore:
\begin{equation*}
\sup_{x\in I}\left\vert R_{4}(x) \right\vert =o\left( \left( \frac{\ln n}{nh}\right) ^{3/4}\right) \text{a.s.}
\end{equation*}
This completes the proof.
\end{proof}

\section*{Acknowledgements}
The first author's research was sponsored by the Spanish FPU grant from MECD with reference FPU13/01371. The work of the first author has been partially carried out during a visit at the Universit\'{e} catholique de Louvain, financed by INDITEX. All the authors acknowledge partial support by the MINECO grant MTM2014-52876-R (EU ERDF support included). The first three authors' research has been partially supported by MICINN Grant MTM2011-22392 (EU ERDF support included) and Xunta de Galicia GRC Grant CN2012/130. The research of the fourth author was supported by IAP Research Network P7/06 of the Belgian State (Belgian Science Policy), and by the contract ``Projet d'Actions de Recherche Concert\'ees'' (ARC) 11/16-039 of the ``Communaut\'e fran\c{c}aise de Belgique'' (granted by the ``Acad\'emie universitaire Louvain''). The authors would like to thank the Associate Editor and the three anonymous re\-fe\-rees for their constructive and helpful comments, which have greatly improved the paper. The authors are grateful to Dr. Sonia P\'ertega and Dr. Salvador Pita, at the University Hospital of A Coru\~{n}a, for providing the colorectal cancer data set.

\section*{References}

\bibliography{mybibfile}

\end{document}